\documentclass[12pt]{aastex631}

\usepackage{graphicx}

\definecolor{purple}{rgb}{1,0,1}

\shorttitle{Very-high-energy Emission from Pulsars}
\shortauthors{Harding et al.}
\graphicspath{{./}{figures/}}

\def\be{\begin{equation}}
\def\ee{\end{equation}}

\def\tt{\mbox{\boldmath $\theta $}}

\def\lsim{\lower 2pt \hbox{$\, \buildrel {\scriptstyle <}\over
         {\scriptstyle \sim}\,$}}

\begin{document}

\title{Very-High-Energy Emission From Pulsars}

\author{Alice K. Harding}
\affiliation{Theoretical Division, Los Alamos National Laboratory, Los Alamos, NM 87545}
\author{Christo Venter}
\affiliation{Centre for Space Research, North-West University, Private Bag X6001, Potchefstroom 2520, South Africa}
\author{Constantinos Kalapotharakos}
\affiliation{University of Maryland, College Park (UMDCP/CRESST), College Park, MD 20742}


\begin{abstract}
Air-Cherenkov telescopes have detected pulsations at energies above 50 GeV from a growing number of \textit{Fermi} pulsars.  These include the Crab, Vela, PSR~B1706$-$44 and Geminga, with the first two having pulsed detections above 1 TeV.  In some cases, there appears to be very-high-energy (VHE) emission that is an extension of the \textit{Fermi} spectra to high energies, while in other cases, additional higher-energy spectral components that require a separate emission mechanism may be present.  We present results of broad-band spectral modeling using global magnetosphere fields and multiple emission mechanisms that include synchro-curvature (SC) and inverse Compton scattered (ICS) radiation from accelerated particles (primaries) and synchrotron-self Compton (SSC) emission from lower-energy pairs.  Our models predict three distinct VHE components: SC from primaries whose high-energy tail can extend to 100 GeV, SSC from pairs that can extend to several TeV and ICS from primary particles accelerated in the current sheet, scattering pair synchrotron radiation, that appears beyond 10 TeV.  Our models suggest that H.E.S.S.-II and MAGIC have detected the high-energy tail of the primary SC component that produces the \textit{Fermi} spectrum in Vela, Geminga and PSR~B1706$-$44.  We argue that the ICS component peaking above 10 TeV from Vela has been seen by H.E.S.S.  Detection of this emission component from the Crab and other pulsars is possible with HAWC and CTA, and directly measures the maximum particle energy in pulsars.  
\end{abstract}



\section{Introduction} \label{sec:intro}
The stunning increase in quality high-energy (HE) pulsar data facilitated by the operation of the \textit{Fermi} Large Area Telescope (LAT; \citealt{Atwood2009}) provided ample opportunity to scrutinise our understanding of the emission mechanisms and magnetospheric geometry responsible for pulsed gamma-ray emission from these enigmatic stellar objects. A generic feature of the GeV spectra observed by the \textit{Fermi} LAT is that they cut off, usually sub-exponentially, in a narrow band around a few GeV. Moreover, a stacking analysis involving 115 \textit{Fermi}-detected pulsars (excluding the Crab pulsar) found no emission above 50~GeV \citep{McCann15}, indicating that one should probably in general not expect detectable levels of very-high-energy (VHE) photons from pulsars. 

Earlier pulsar models indeed predicted detectable TeV fluxes from these sources, subject to some modelling uncertainties and approximations. For example, simulations invoking the standard outer gap (OG) model yielded spectral components beyond 100~GeV for some pulsars when estimating the inverse Compton scattering (ICS) flux of primary electrons on synchrotron radiation (SR) by secondary pairs \citep{Cheng1986I,Cheng1986II,Romani1996}. This resulted in a natural bump around a few TeV that reached $\lesssim5\%$ of the GeV flux \citep{Hirotani2001,Takata2006}, depending on model assumptions. However, it was not clear if such components would survive magnetic one-photon or two-photon absorption, so their detectability remained uncertain.  Most recently \citet{Rudak2017} used an OG model to predict ICS emission for Vela above 10 TeV from primary gap-accelerated particles scattering optical/IR emission from secondary pairs. Some polar cap (PC) or pair-starved PC models predicted curvature radiation (CR) spectral cutoffs of up to 100 GeV \citep{Bulik2000,Harding2002a}. 
Later such models  \citep{Harding2005,Venter2005,Frackowiak2005}, a slot gap (SG) model \citep{Harding2008} and an OG model \citep{Tang2008} predicted CR cutoffs around $\sim10-50$~GeV (still prior to the launch of the \textit{Fermi} LAT, though). One could conclude that this CR component may just fall short of the reach of ground-based Cherenkov detectors, the energy threshold of which exceeded 100~GeV.  

Given the uncertain and low expectations for pulsed TeV spectral components by some models and observations, it was thus surprising that MAGIC announced the detection of pulsations from the Crab pulsar at energies up to $\sim25$~GeV \citep{Aliu2008,Aleksic2011,Aleksic2012}, followed by the VERITAS detection of pulsed photons up to $\sim400$ GeV \citep{Aliu2011}, and finally, MAGIC's detection of pulsations up to $1.5$~TeV \citep{Ansoldi2016}. This opened a new window into pulsar science. H.E.S.S.-II next detected pulsed emission from the Vela pulsar in the sub-20 GeV to 100~GeV range \citep{Abdalla2018}. The latest observations by H.E.S.S.\ unveil pulsed emission from this pulsar at a few TeV (H.E.S.S.\ Collaboration, in preparation). Additionally, H.E.S.S.-II detected pulsed emission from PSR~B1706$-$44 in the sub-100~GeV energy range \citep{SpirJacob2019}. MAGIC recently detected pulsed emission from the Geminga pulsar between 15~GeV and 75~GeV at a significance of 6.3$\sigma$ \citep{Acciari2020}. However, only the second light curve peak is visible at these energies. The measured spectrum is smoothly connected to that measured by the \emph{Fermi} LAT, and may perhaps indicate a transition from a CR to ICS spectral component.

Some trends are becoming apparent. Perhaps the most striking is the continuation of the decrease in the ratio of first to second light curve peaks  \citep{Abdo2010PC1,Abdo2013} toward higher energies: as the photon energy $E_{\gamma}$ is increased beyond several GeV, the intensity ratio of the first (P1) to second (P2) light curve peak decreases for Vela and Geminga. In the GeV band, we attribute this to the systematically larger local curvature radius, $\rho_{\rm c}$, of trajectories of particles responsible for the second peak, compared to the first, leading to a relatively larger spectral cutoff for P2 (Barnard et al., submitted). The highest-energy particles may also be responsible for this effect in the TeV band via ICS emission. Furthermore, one notices that the $\gamma$-ray peaks of Crab, Vela and Geminga remain at the same phase positions, that the inter-peak (bridge) emission evolves in the case of Vela, and that the peak widths decrease for Crab \citep{Aliu2011}, Vela \citep{Abdo2010Vela} and Geminga \citep{Abdo2010Geminga} at increased photon energy. 

In this paper, we attempt to explain the new subTeV phenomenology and also make predictions of additional VHE emission components at higher energy to guide further observations by ground-based facilities. We model various spectral components, taking into account several recent developments, such as force-free (FF; \citealt{Kalapotharakos2009}) magnetic structures, and also indications from the FF inside and dissipative outside (FIDO) models \citep{Brambilla2015,Kalapotharakos2014,Kalapotharakos2017} and particle-in-cell simulations (PIC; \citealt{Brambilla2018,Cerutti2016current,Cerutti2016,Cerutti2020,Kalapotharakos2018,Philippov2018}) that predominant dissipation is happening in the current sheet, rather than interior to the light cylinder. The added data from the TeV window may play an important role in constraining particle energetics and break degeneracies that exist when only considering lower-energy data. 

The structure of the rest of the paper is as follows. In Section~\ref{sec:model}, we give a brief description of our model and calculation of the broadband spectra. In Section~\ref{sec:results}, we present phase-averaged and phase-resolved spectra for several pulsars: Vela, Crab, Geminga, PSR B1706-44 and PSR J0218+4232. Our discussion and conclusions  follow in Section~\ref{sec:disc}.

\section{Description of Model}\label{sec:model}

Much of the magnetic field geometry, particle trajectory and radiation calculations are the same as in \citet{Harding2018}, (hereafter H18), which is an updated version of the model used in \citet{Harding2015} (hereafter HK15).  Since the details may be found in those papers, we will give only a summary of the main elements of the model and detailed descriptions of only what is new to the present calculations.  

There has been recent rapid progress in understanding the structure, current flow and particle acceleration in global pulsar magnetospheres.  Simulations of FF (where $\mathbf{E} \cdot \mathbf{B} = 0$ and $\mathbf{E} < \mathbf{B}$ is assumed) and dissipative magnetospheres \citep{Kalapotharakos2012,Li2012}, where a macroscopic conductivity is assumed, have shown that reproduction of \textit{Fermi} $\gamma$-ray pulsar light curves requires near-FF magnetospheres \citep{Kalapotharakos2014,Kalapotharakos2017}.  The region of highest accelerating electric field components are found near the current sheet outside the light cylinder at a radius $R_{\rm LC} = c/\Omega$, where $\Omega$ is the pulsar angular rotation rate and $c$ is the speed of light, indicating that most of the particle acceleration takes place there.  More recent kinetic or PIC simulations of global magnetospheres \citep{Kalapotharakos2018,Philippov2018} confirm this finding by showing that the highest-energy particles are found in the current sheet.
In our calculations, we adopt the 3D magnetic field geometry of an FF magnetosphere \citep{Kalapotharakos2012} in which we inject both particles that enter the current sheet and are accelerated to high energy (primaries), and electron-positron pairs assumed to come from pair cascades near the NS surface (see \S \ref{sec:pair}).  All particles are injected at foot points of open field lines at the PC (see \S \ref{sec:mag}) and radiate synchro-curvature (SC) and inverse-Compton  scattering (ICS) emission (see \S \ref{sec:spec}) along their trajectories in the inertial observer frame (IOF).  The radiated photons from each type of particle, pairs and primaries, are collected in arrays of energy, observer angle $\zeta$ with respect to the rotation axis and rotation phase $\phi$, for a fixed magnetic inclination angle $\alpha$.  Sky maps of {the intensity distribution in} $\zeta$ vs. $\phi$ can then be constructed in different energy ranges and from these we can make energy-dependent light curves and phase-resolved spectra.

\subsection{Magnetosphere geometry} \label{sec:mag}

The FF magnetosphere magnetic fields for various magnetic inclination angles, $\alpha$, are generated in Cartesian grids with resolution $0.02\, R_{\rm LC}$, from the NS ``surface" at $0.2\, R_{\rm LC}$ to $2.0\, R_{\rm LC}$.  To compute particle trajectories and radiation, the local magnetic field values are obtained by 3D interpolation of the Cartesian grid.  Since the actual NS surface is much smaller than the lower limit of the grid, we extend the field values to the real pulsar surface by a ramp function, joining the numerical FF solution with an analytic retarded vacuum dipole \citep{Deutsch1955,Dyks2004}.

To compute the particle trajectories, we first determine the PC rims by finding the last open field lines at each magnetic azimuth by fourth-order Runge-Kutta integration \citep{Dyks2004} and the open volume coordinates (ovc) that are the radial, $r_{\rm ovc}$, and azimuthal, $l_{\rm ovc}$, divisions of the distorted PCs.  Particles are then injected on the neutron star (NS) surface at particular $(r_{\rm ovc}, l_{\rm ovc})$ points, where $r_{\rm ovc} = 0$ labels the magnetic axis and $r_{\rm ovc} = 1$ labels the PC rim.  For the calculations in this paper we use between 360 and 720 azimuthal divisions and inject pairs from $r_{\rm ovc} = 0.8$ to $r_{\rm ovc} = 0.9$ and primaries from $r_{\rm ovc} = 0.9$ to $r_{\rm ovc} = 0.96$.  For the Crab, the pairs are injected between $r_{\rm ovc} = 0.88$ and $r_{\rm ovc} = 0.91$ and the primaries between $r_{\rm ovc} = 0.91$ and $r_{\rm ovc} = 0.94$.  We have observed in our PIC models \citep{Kalapotharakos2018} that the width of the acceleration region in the current sheet grows smaller with increasing pair injection.  Since very energetic pulsars like the Crab have larger pair multiplicity, narrower widths over which particle are injected and accelerated are expected.  Primary particles are injected with low Lorentz factors $\gamma_p = 200$ and pairs are injected with a spectrum of Lorentz factors (see \S \ref{sec:pair}).  

We compute the particle trajectories in the IOF as in HK15, with the velocity being a sum of a component parallel to the magnetic field and the local drift velocity.  The dominance of the drift velocity near and beyond the light cylinder allows the particle velocity to remain subluminal while drifting backward (in phase) along field lines.  However, in the present calculation the particle trajectories and their radii of curvature, assumed to be independent of particle energy, are first computed using interpolation and smoothing, with fine spatial resolution and stored in tables.  The subsequent calculation of the particle dynamics and radiation at different Lorentz factors that use dynamic step sizes then interpolate from these pre-computed tables.  This method of computing trajectories, detailed in Barnard et al., in prep., gives a more accurate determination of the trajectory radii of curvature.

\subsection{PC pair spectra} \label{sec:pair}

Electron-positron pair cascades at the PCs are necessary to supply plasma to the magnetosphere, since extraction of charges of a single sign induce large accelerating electric fields, $E_\parallel$, and cannot produce the current required by the global solution \citep{Timokhin2010,Timokhin2013} over the whole PC.  Pairs are also necessary to screen $E_\parallel$ throughout most of the magnetosphere to produce the near-FF global conditions.  The cascades are initiated by high-energy photons that are produced by particles accelerated in the induced $E_\parallel$, creating pairs by the one-photon process in the strong magnetic field.  The pairs produce SR photons that create multiple (up to 5) generations of pairs \citep{Daugherty1982}. \citet{Timokhin2010} and \citet{Timokhin2013} found that the need to supply the global, near-FF current as a boundary condition produces non-steady pair cascades, where pairs are created in bursts that completely screen the $E_\parallel$, which reappears after the pairs exit the gap.  However, the multiplicity of pairs created to screen the gap is small in comparison to the multiplicity of the full cascade, most of which occurs above the gap \citep{Timokhin2015}.  The region of particle acceleration and screening is also much smaller than the typical distance over which the full cascade extends.  We can therefore calculate the spectrum of pairs from the full cascade using the energy of the gap-accelerated particles.

To generate the pair spectra for the calculations of this paper, we use the estimate of gap particle Lorentz factor from  Eqn (23) of \citet{Timokhin2019} 
\be
\gamma_{\rm gap} \simeq 3.2 \times 10^7\,P^{-1/7}\,B_{12}^{-1/7}\,\rho_{c,7}^{4/7},
\ee
which is similar to the expression derived by \citet{Harding1998} for steady gaps, where $P$ is the pulsar period in s., $B_{12}$ is the surface magnetic field strength in units of $10^{12}$ G, and $\rho_{c,7}$ is the field line radius of curvature at the surface in units of $10^7$ cm.  The major characteristic is that the gap particle energy is very insensitive to surface magnetic field and period but is most sensitive to field line radius of curvature.  Although there is a distribution of particles moving upward from the gap, the highest-energy particles that produce most of the pairs are nearly monoenergetic. 
The results of \citet{Timokhin2015} adopted a 1D gap that did not take into account the variation of curvature radius across the PC which causes a variation in accelerated particle energy \citep{Harding1998}. 
We use the estimate above of the gap particle energy as the initial energy for the full cascade and compute the pair spectra from the whole PC, as in \citet{Harding2011}, varying $\rho_c$ which determines initial energy.  Although we have an estimate for the initial cascade particle energy the multiplicity, the ratio of the density of high-energy particles from the gap to the Goldreich-Julian density, $\rho_{\rm GJ} = (B\Omega/\pi ec)$ is not as well determined.  We will therefore take the pair cascade multiplicity $M_+$ (number of pairs created via the seed emission of a single primary) as a variable parameter.  The pair spectrum, however, is determined by the initial particle energy and the magnetic field geometry.

\begin{figure} 
\includegraphics[width=200mm]{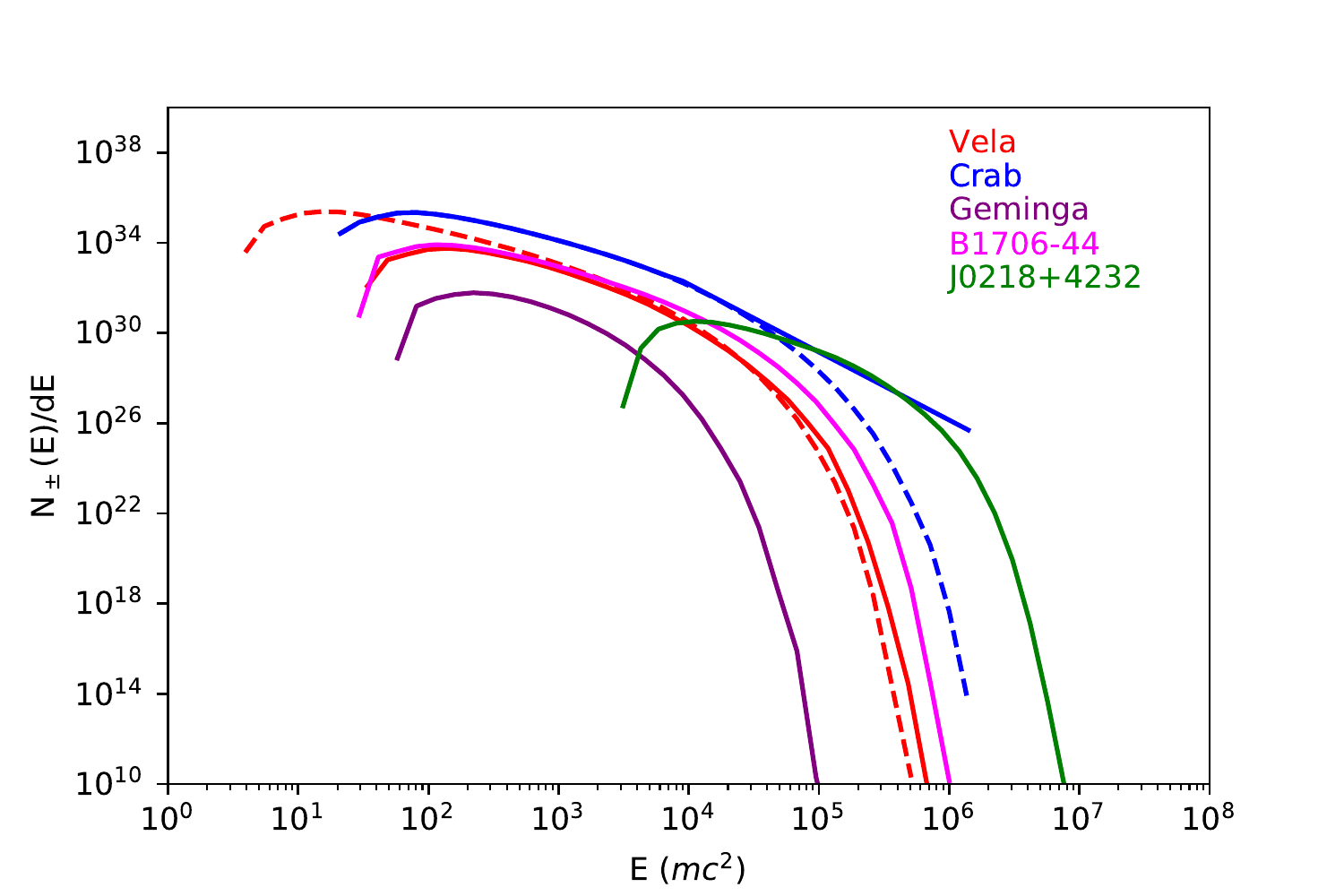}
\caption{Electron-positron pair spectra for the sources modeled in this paper, from a polar cap cascade simulation.  The solid curves for Vela,  B1706-44 and Geminga, and the dashed curve for the Crab assume centered dipole magnetic fields.  The dashed curve for Vela assumes a non-dipolar surface field with offset parameter $\epsilon = 0.4$ \citep{Harding2011}, while the J0218+4232 spectrum assumes a non-dipolar field with $\epsilon = 0.6$.  The solid curve for the Crab is the dipole pair spectrum with a power-law extension.}
\label{fig:pair}
\end{figure}

Figure \ref{fig:pair} shows the pair spectra for the sources we model in this paper.  The different pulsars have a variety of spectral shapes, with very energetic pulsars like the Crab displaying the largest range of pair energies extending from Lorentz factors of 20 to above $10^6$.   Middle-aged pulsars like Vela and B1706$-$44 have pair spectra that cut off at somewhat lower energies, and older pulsars like Geminga have spectra with an even more limited energy range.  MSPs such as J0218+4232 have pair spectra with much higher low and high-energy cutoffs, ranging from $\gamma \sim 5000$ to nearly $10^7$.   Since they have much lower surface field strengths compared with young pulsars, the photons must have much higher energies to reach the threshold for pair production, leading to higher pair energies.  The spectra with solid lines in Figure \ref{fig:pair} for Vela, B1706$-$44, Geminga and J0218+4232 and the dashed-line spectrum for the Crab are products of pair cascade models that assumes a pure dipole magnetic field.  The dashed-line spectrum for Vela and the solid-line spectrum for the Crab are alternative models we explore.  The Vela dashed-line spectrum is a product of a pair cascade model that assumes a non-dipolar field structure in which a toroidal perturbation has been added to a dipole \citep{Harding2011}, parametrized by the quantity $\epsilon$.   The Vela pair spectrum shown in Figure \ref{fig:pair} assumes $\epsilon = 0.4$.  The alternative Crab pair spectrum assumes a power-law extension of the dipole pair spectrum, which would be necessary to account for the COMPTEL soft $\gamma$-ray points (see \S \ref{sec:Crab}).

It was found that the pair cascades that match the global currents can only operate in PC regions where the current $J/J_{\rm GJ} >1$ or $J/J_{\rm GJ} < 0$ \citep{Timokhin2013}, where $J_{\rm GJ} = \rho_{\rm GJ}\, c$.    Therefore, in addition to the pair injection regions that are limited by the $r_{\rm ovc}$ range, we further limit the pair injection to the regions where the global return current $J/J_{\rm GJ} < 0$ and connect to the current sheet (by limiting injection to a range of $l_{\rm ovc} = 1.5 - 4.78$ radians).  For oblique rotators, this region is located on field lines orientated toward the spin equator.  

\subsection{Calculation of broadband photon spectra} \label{sec:spec}

Our modeling of the broadband radiation of the electron-positron pairs and accelerated particles (primaries) follows closely to that of HK15 and details of the calculations can be found there.  There are a number of differences in the present calculation that we will detail below.  
The primaries and pairs are injected at the NS surface as described in previous sections.  The primaries are accelerated by $E_\parallel$ fields along their trajectories with assumed values that are lower below the light cylinder and higher in the current sheet.  This two-tier $E_\parallel$ is guided by PIC models that show the highest acceleration in the current sheet starting at the so-called Y-point (i.e., the tip of the closed field line zone) and a low level of acceleration below the light cylinder \citep{Brambilla2018}.  The pairs are injected with a range of energies and relative fluxes equal to those of the pair spectra shown in Figure \ref{fig:pair} and are not accelerated along their trajectories.  As in HK15, the code loops through all particle trajectories twice, the first time to compute the SC radiation and 3D grid of emissivities and the second time to compute the ICS emission using the SC emissivities from the first loop.  Also as in HK15, we assume that both pairs and primaries are injected with zero pitch angle but can acquire finite pitch angles through resonant absorption of radio photons that are emitted at a specified altitude, $r_{\rm radio}$, above the NS surface. However, the observed phase alignment of the light curves from radio to TeV $\gamma$-rays for the Crab implies that the radio emission also comes from the current sheet \citep{Philippov2019}.  The cone beam geometry used in the radio absorption calculation of HK15 cannot be applied outside the light cylinder.  Therefore, in our model for the Crab, to simulate absorption of radio emission in the current sheet, we give the particles a constant pitch angle above a radius of $0.8\,R_{\rm LC}$, with pairs maintaining a pitch angle $\psi = 2 \times 10^{-3}$ and primaries $\psi = 10^{-5}$.  

The primaries will therefore radiate a mixture of SR and CR that is captured by treating the radiation as SC, with the emission below the light cylinder being mostly SR, and becoming mostly CR in the current sheet as particle Lorentz factors exceed $10^7$.  Since the pairs have relatively low Lorentz factors, their radiation in the first loop is in the SR-limit of SC emission.  We use the SC expression for the radiated power per photon energy interval presented in \citet{Torres2018} 
\be
{dP_{\rm SC} \over dE} = {\sqrt{3} e^2 \gamma y \over 4\pi \hbar r_{\rm eff}}\, \left[(1+z)\,F(y) - (1-z)\,K_{2/3}(y)\right],
\ee
where
\be
y = \varepsilon/\varepsilon_c,  ~~~~~z = (Q_2\,r_{\rm eff})^{-2},
\ee
\be
\varepsilon_c = \frac{3}{2} \hbar c Q_2 \gamma^3
\ee
\be
F(y) = \int_y^\infty\,K_{5/3}(y) dy,
\ee
\be
r_{\rm eff} = {\rho_c \over \cos^2\psi}\,\left[ 1 + \xi + {r_g \over \rho_c} \right]^{-1}
\ee
is the effective curvature radius,
\be
r_g = {mc^2\gamma \,\sin\psi \over eB}
\ee
is the gyroradius, $\psi$ is the pitch angle,
\be
Q_2 = {\cos^2\psi \over \rho_c}\,\left(1+3\xi+\xi^2+{r_g\over \rho_c} \right)^{1/2},
\ee
and
\be
\xi = {\rho_c \sin^2\psi \over r_g \cos^2\psi}.
\ee
The parameter $\xi$ determines the dominant SC regime, with SR dominating when $\xi \ll 1$ and CR when $\xi \gg 1$.

To compute the ICS emission of both pairs and primaries, the photon density at each particle position along its trajectory is calculated from the stored SC emissivities from the first loop, as detailed in HK15.  The particle dynamics follows the Lorentz factor and perpendicular momentum, $p_{\perp}$, along each trajectory, integrating Eqns (39) and (40) of HK15, with the only difference being the replacement of separate SR and CR loss rates with the SC loss rate
\be
{d\gamma \over dt} = {2e^2\,\gamma^4\,c \over 3\rho^2_c mc^2}\, g_r,
\ee
where
\be
g_r = \left({\rho_c \over r_{\rm eff}}\right)^2\,{[1+7(r_{\rm eff}Q_2)^{-2}] \over 8(Q_2 r_{\rm eff})^{-1}}
\ee
is a factor that determines the dominance of CR ($g_r \sim 1$) or SR losses ($g_r \gg 1$).

In the first trajectory loop, the ICS losses are zero so only SC losses and $E_\parallel$ acceleration (for primaries) are at play.  In the second trajectory loop, SC losses are included in the particle dynamics since they dominate over ICS losses, justifying the neglect of the latter in the first loop. Also in the second loop, we do not need to compute the perpendicular particle momentum $p_{\perp}$ along the trajectory, since the ICS emission depends only on $\gamma$.

In this calculation we introduce attenuation of the VHE emission due to $\gamma$-$\gamma$ pair production.  Our approximate treatment is to multiply the ICS flux by an attenuation factor
\be
\dot N_{\rm IC} = \dot N_{\rm IC,0}\, \exp[-\tau_{\gamma\gamma} (\varepsilon)]
\ee
with
\be \label{eq:tau}
\tau_{\gamma\gamma} = \int_0^s \, n_\gamma\,\sigma_{\gamma\gamma}\,ds' \sim n_\gamma\,\frac{\sigma_{\rm T}}{3}\,\Delta s,
\ee
with $\sigma_{\rm T}$ the Thomson cross section.
The VHE photons of energy $\varepsilon$ scatter primarily with photons of energy
\be
\varepsilon_{\rm s} = \frac{2}{\varepsilon_\gamma (1-\cos\theta)} \simeq \frac{4\,
\gamma_{\rm x}^2}{\varepsilon_\gamma},
\ee
since $\theta \simeq 1/\gamma_{\rm x}$, where $\gamma_{\rm x}$ is the Lorentz factor of the Optical/IR emitting pairs which we set to 10.
In Eqn (\ref{eq:tau}), $n_\gamma$ is the soft-photon density which we approximate as 
\be
 n_\gamma \simeq \frac{L_s}{\varepsilon_{\rm s} A c}
\ee
where $L_s$ is the luminosity at energy $\varepsilon_{\rm s}$ and $A \simeq 0.1\,R_{\rm LC}^2$ is the approximate area.  Also in Eqn (\ref{eq:tau}) we set 
$\Delta s \sim (2\,R_{\rm LC} - r_{\rm p})$, where $r_{\rm p}$ is the position of the ICS-emitting particle, so that $\Delta s$ is the path length that the VHE photon will travel to exit the computation box.  

By estimating the mean-free path $\ell_{\gamma\gamma} \simeq 1/n_\gamma \sigma_{\gamma\gamma}$, the Crab has  $\ell_{\gamma\gamma} \sim 4 \times 10^6 - 1.5 \times 10^7$ cm $< R_{\rm LC}$, while Vela has $\ell_{\gamma\gamma} \sim 10^{10}$ cm $\gg R_{\rm LC}$.  The other sources have mean-free paths similar to or larger than Vela's.  Therefore we find that $\gamma$-$\gamma$ attenuation will only be important for the Crab.

The major differences, then, between the model in this paper and that of HK15 are:
1) The SC radiation spectrum and energy losses are implemented for primary particles.
2) The SC emissivity does not cover the full computational grid but has spatial upper limits that are tailored to the extent of the soft emission for each pulsar to provide better resolution. For Vela, Geminga and B1706-44 the emissivity grid extends to $0.8\,R_{\rm LC}$ while for the Crab and J0218+4232 it extends to $2.0\,R_{\rm LC}$.
3) The spectral energy range has been extended to cover 18 decades, from IR to 100 TeV.
4) The computation of particle trajectories and their radii of curvature is more accurate.
5) The $E_\parallel$  now has different constant values inside and outside the light cylinder, whereas HK15 assumed the same value from the NS surface to $2\,R_{\rm LC}$.
6) The pair injection occurs only over part of the PC, where $J/J_{\rm GJ} < 0$.

All of the above changes were implemented in H18.  The model in this paper has the additional changes that were not in HK18:
1) We corrected an error of factor $\sim2$ in the ICS radiation, where the energies in $mc^2$ were not converted to MeV.
2) We updated the radio luminosity values for Vela.
3) $\gamma$-$\gamma$ pair attenuation of the ICS radiation is now included.

Table 1 lists both the observed and the model parameters for the sources we study in this paper.  All of the radio fluxes listed are those measured at 400 MHz, except for Geminga which has no measured radio emission.  For Geminga we therefore assume that it produces radio emission that is not beamed in our direction and assume a value of 1000 mJy at 400 MHz to model the pair SR.  The parameters that are varied to match  the observed SEDs and high-energy light curves are the magnetic inclination angle, $\alpha$, the observer angle $\zeta$, the low and high values of accelerating electric field, $R_{\rm acc}^{low} = eE_\parallel^{low}/mc^2$ and $R_{\rm acc}^{high} = eE_\parallel^{high}/mc^2$, the primary current, $J/J_{\rm GJ}$, the pair multiplicity, $M_+$ and the radio emission altitude, $r_{\rm radio}$. We do not consider the viewing angle $\zeta$ as a free parameter since, for a given $\alpha$, it is adjusted to match the {\sl Fermi} light curves.

\begin{table}
\caption{Source and Model Parameters}
\vskip 0.5 cm
\hskip -2.0cm
\begin{tabular}{cccccccccc}
\hline\noalign{\smallskip}

Pulsar & $P$ & $d$ & Radio flux & $\alpha$ & $R_{\rm acc}^{low}$ & $R_{\rm acc}^{high}$ & $J/J_{\rm GJ}$ & $M_+$ & $r_{\rm radio}$  \\
      & (s) & (kpc) & (mJy) & & $(\rm cm^{-1})$ & $(\rm cm^{-1})$ &  & & ($R_{\rm LC}$) \\
\hline
Vela & 0.089 & 0.25 & 5000 &$75^\circ$ & 0.04 & 0.2 &18 & $6 \times 10^3$ & 0.1/0.2\\
Crab & 0.033 & 2.0 & 700 &$45^\circ$  &  0.04 & 0.4 & 5.0 & $3 \times 10^5$ & $>$ 0.8 \\
B1706$-$44 & 0.102 & 2.3 & 25 & $45^\circ$/$30^\circ$ & 0.04 & 0.2 & 20 & $6 \times 10^4$ & 0.08\\
Geminga & 0.237 & 0.25 & 1000 & $75^\circ$ & 0.04 & 0.15 & 10 & $2 \times 10^4$ & 0.1 \\
J0218+4232 & 0.0023 & 3.1 & 100 & $60^\circ$ & 0.5 & 5.0 & 150 & $3 \times 10^5$ & 0.38\\
\hline
\end{tabular}
\end{table}

\section{Results}\label{sec:results}

Here we discuss the results of our modeling, first presenting the calculated phase-averaged spectral energy distributions (SEDs) for all sources, followed by the energy-dependent sky maps and light curves.  For all sources, the $E_\parallel^{low}$, $E_\parallel^{high}$ and $J/J_{\rm GJ}$ parameters are adjusted so that the SC component matches the hard X-ray to GeV data, particularly the high-energy cutoff in the \textit{Fermi} SEDs which is especially sensitive to $E_\parallel^{high}$.  The hard X-ray SED is most sensitive to the value of $E_\parallel^{low}$.  The $M_+$ and $r_{\rm radio}$ are adjusted so that the pair SR matches the observed IR to soft X-ray data.  The viewing angle, $\zeta$, is chosen to best match the observed \textit{Fermi} and VHE light curves.  After these parameters are adjusted, the ICS component fluxes and SEDs are determined without further parameter adjustments (the ICS intensity thus being anchored by the lower-energy components).

\subsection{Vela pulsar} \label{sec:Vela}

\begin{figure} 
\includegraphics[width=180mm]{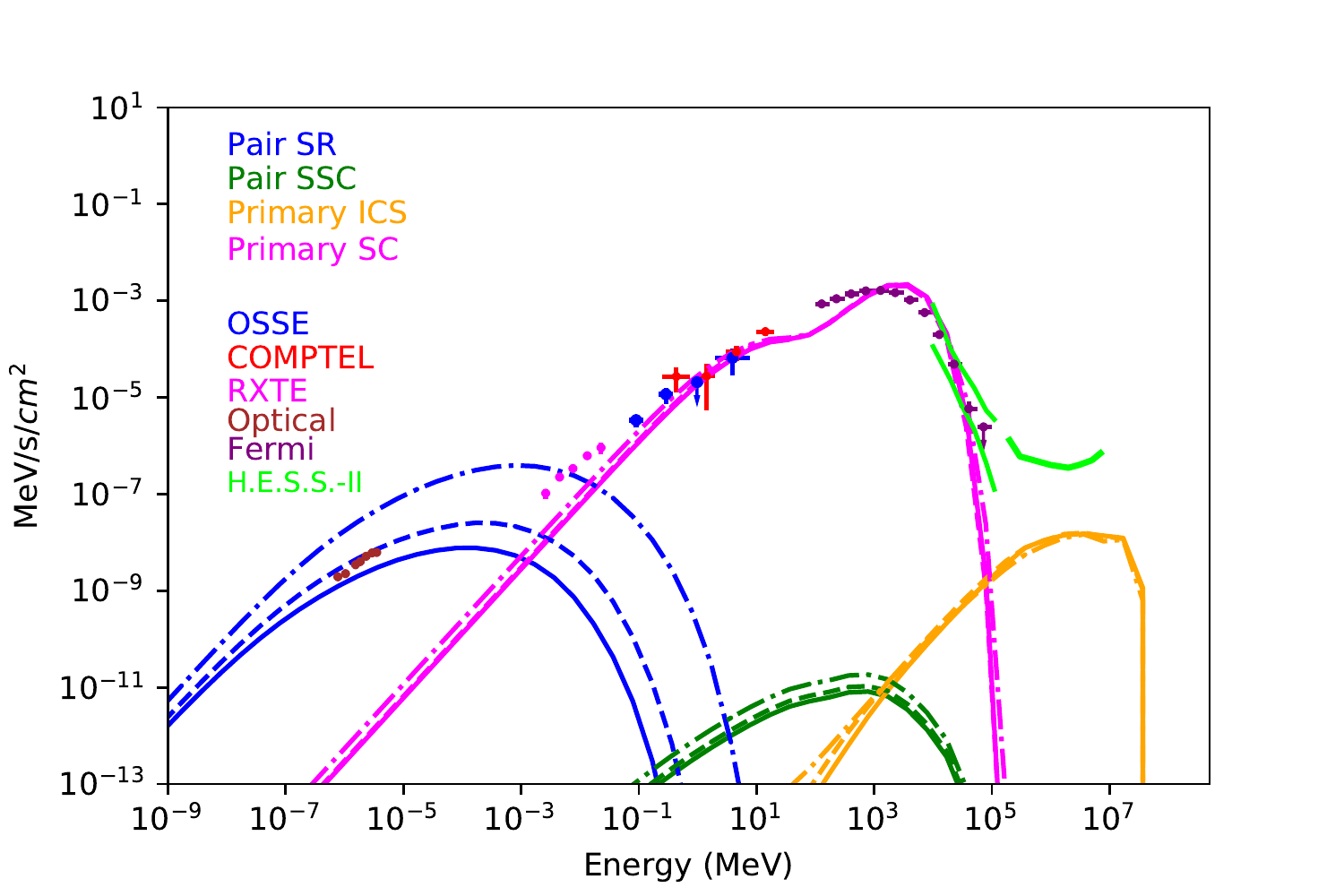}
\caption{Model SED for the Vela pulsar for inclination angle $\alpha = 75^\circ$ and three different viewing angles $\zeta =  40^\circ$ (solid), $43^\circ$ (dashed) and $50^\circ$ (dot-dashed), using the solid pair spectrum (dipole field) shown in Figure \ref{fig:pair} with $M_+ = 6 \times 10^3$.  Data points are from \citet{Abdo2013} (http://fermi.gsfc.nasa.gov/ssc/data/access/lat/2nd\_PSR\_catalog/), \citet{Shibanov2003}, and \citet{Harding2002}. The H.E.S.S.-II detection \citep{Abdalla2018} and high-energy sensitivity for 50 hr of observation \citep{Holler2015} are also shown.}  
\label{fig:Vela_spec0}
\end{figure}

Figure \ref{fig:Vela_spec0} shows a model SED for the Vela pulsar using the solid pair spectrum in Figure \ref{fig:pair}, for three values of the viewing angle.  This result is very similar to the model in H18 who used the same pair spectrum for a pure dipole and the same $E_\parallel^{low}$, $E_\parallel^{high}$ and $M_+$ values.  While the pair SR spectra are sensitive to viewing angle, the high-energy primary SC and ICS spectra are much less sensitive.  The primary ICS flux primarily depends on the low-energy part of the pair SR spectrum which the high-energy primaries can scatter in the Thompson limit, while most of the sensitivity to $\zeta$ in the pair SR spectra is at the highest energies.  In the soft X-ray range, the observed Vela spectrum is dominated by a strong thermal component which we have not plotted.  There is a mismatch of the primary SC component with both the RXTE hard X-ray spectrum and the low energy part of the {\sl Fermi} spectrum around 100 MeV.  This is a consequence of our simplified two-valued $E_\parallel$ distribution that creates an artificial dip in the SC spectrum where the $E_\parallel$ transition occurs at the light cylinder.  A continuously variable $E_\parallel$ distribution that started at lower values and had no sharp jump at the light cylinder would produce a higher level of hard X-ray emission and a smoother  transition to the {\sl Fermi} spectrum.  Future modeling could use the $E_\parallel$ distribution from dissipative or PIC global models as a guide. The shape of the primary SC spectrum determines the shape of the primary ICS spectrum below the cutoff (where the photon energy equals the maximum particle energy).  Consequently, filling in the dip in the SC spectrum by a variable $E_\parallel$ to match the low-energy {\sl Fermi} spectrum would raise the flux of the primary ICS spectrum. 

\begin{figure} 
\includegraphics[width=180mm]{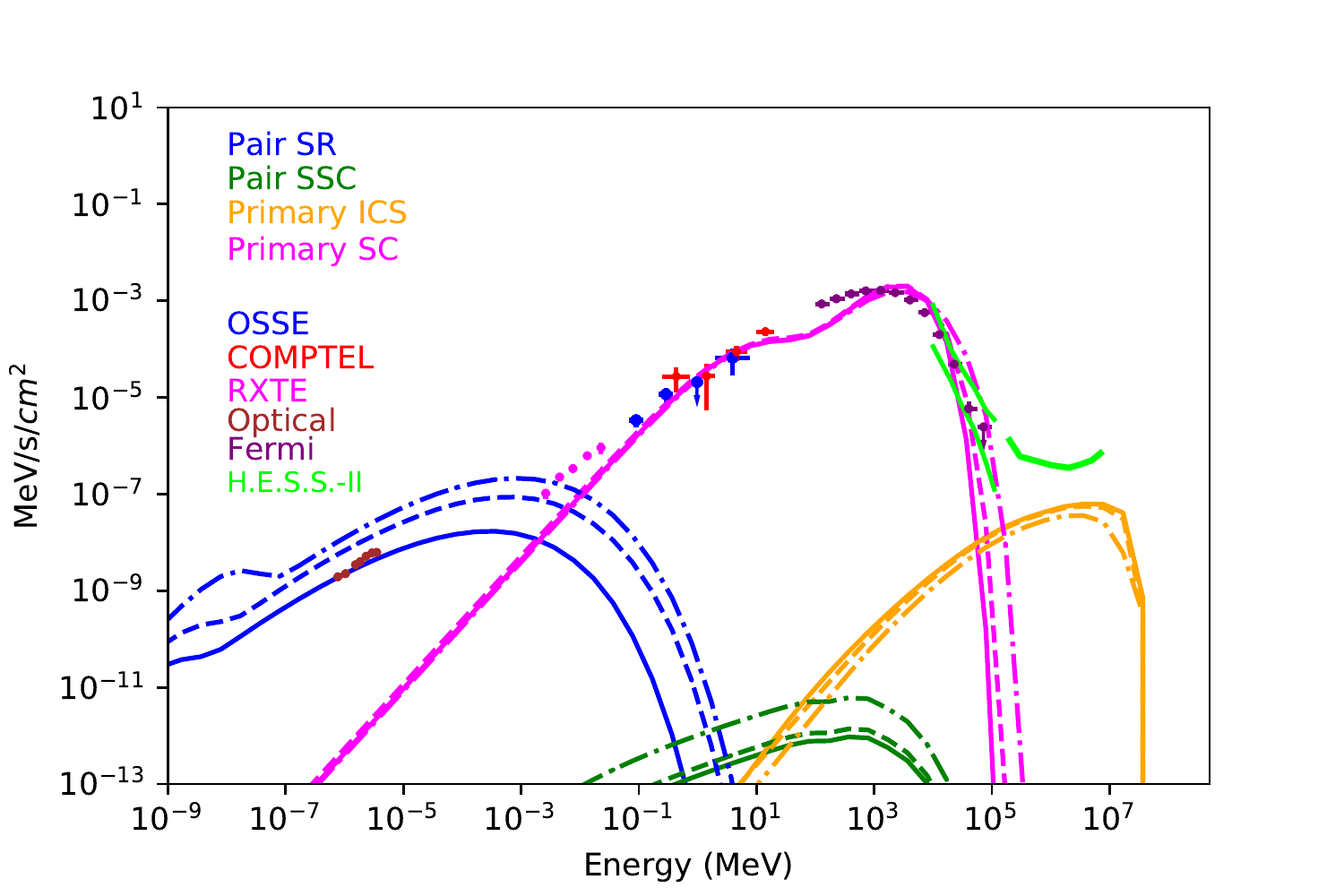}
\caption{Model SED for the Vela pulsar for inclination angle $\alpha = 75^\circ$ and three different viewing angles $\zeta =  46^\circ$ (solid), $50^\circ$ (dashed) and $70^\circ$ (dot-dashed), using the dashed pair spectrum (non-dipole field) shown in Figure \ref{fig:pair} with $M_+ = 6 \times 10^3$}. 
\label{fig:Vela_spec1}
\end{figure}

Figure \ref{fig:Vela_spec1} shows a model SED using instead the dashed pair spectrum in Figure \ref{fig:pair}, for three values of the viewing angle.  The result is similar to that shown in H18 but with notable differences.  This pair spectrum assumes a non-dipolar field but the same $M_+$.  Since the non-dipolar pair spectrum extends to lower pair energy, the pair SR spectrum also extends to lower photon energy and the IR/Optical flux is higher.  Consequentially, this model's primary ICS flux is higher by almost an order of magnitude and is closer to the H.E.S.S.-II threshold.  The primary SC component is the same as shown in Figure \ref{fig:Vela_spec0}  since the $E_\parallel^{low}$ and $E_\parallel^{high}$ values are the same.  Viewing angles in the range $50^\circ$ to $70^\circ$ make minor differences to the primary ICS component.

\begin{figure} 
\includegraphics[width=180mm]{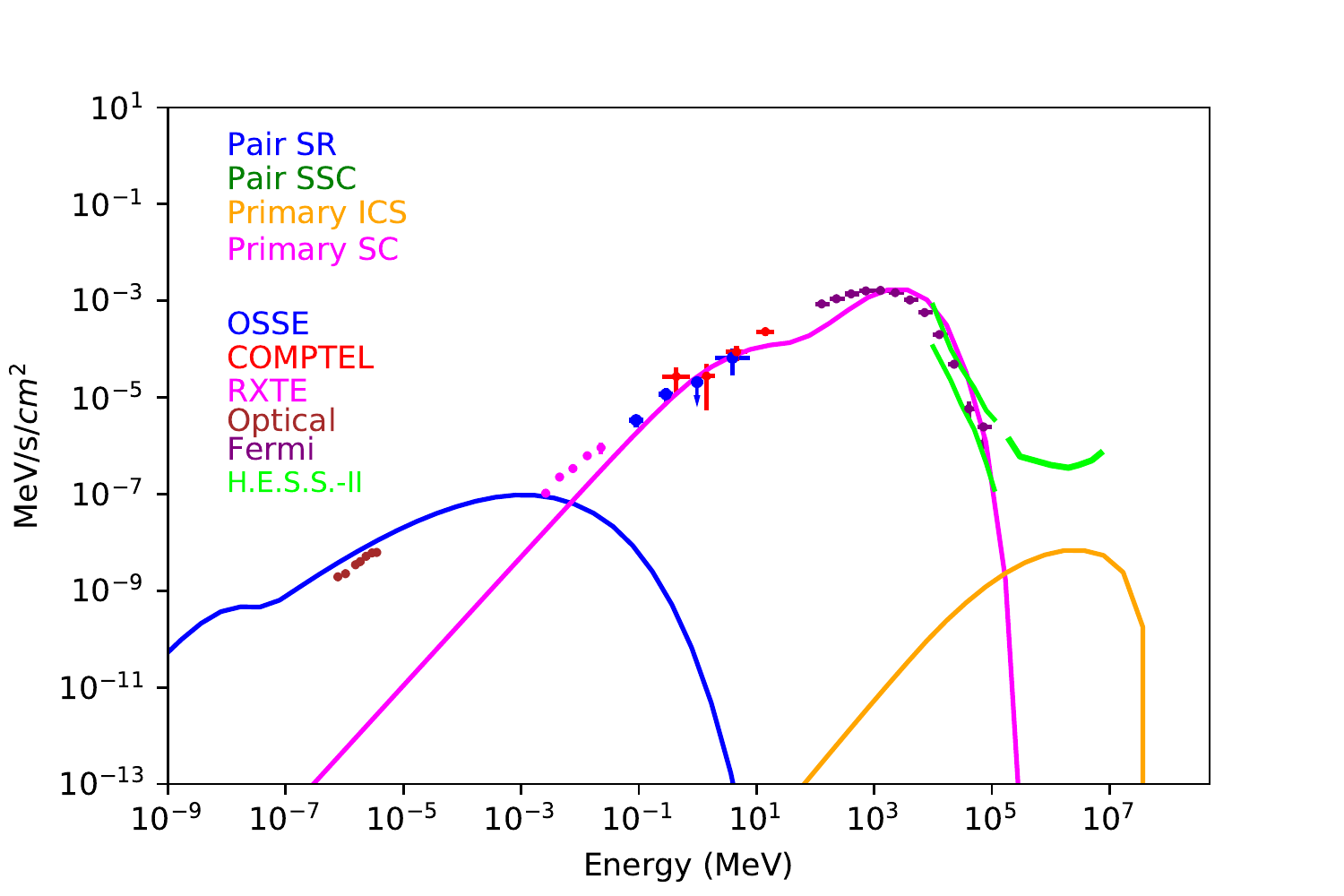}
\caption{Model SED for the Vela pulsar for inclination angle $\alpha = 75^\circ$ and viewing angle $\zeta =  60^\circ$, using the dashed pair spectrum (non-dipole field) shown in Figure \ref{fig:pair} with $M_+ = 1 \times 10^3$.}  
\label{fig:Vela_spec2}
\end{figure}

Figure \ref{fig:Vela_spec2} shows a model SED for the same parameters and pair spectrum, but with a lower pair multiplicity, $M_+ = 1 \times 10^3$.  In this case, the primary ICS flux is about a factor of ten lower and closer to that predicted by H18, even though the pair SR spectrum here has a higher optical/IR flux.  Although the pair SR spectrum for the $\zeta = 46^\circ$ case in Figure \ref{fig:Vela_spec1} matches the observed optical spectrum as well, the higher values of $\zeta$ agree better with the $\zeta \sim 64^\circ$ measured from the pulsar wind nebula torus image fits \citep{Ng2008}.

\begin{figure} 
\includegraphics[width=200mm]{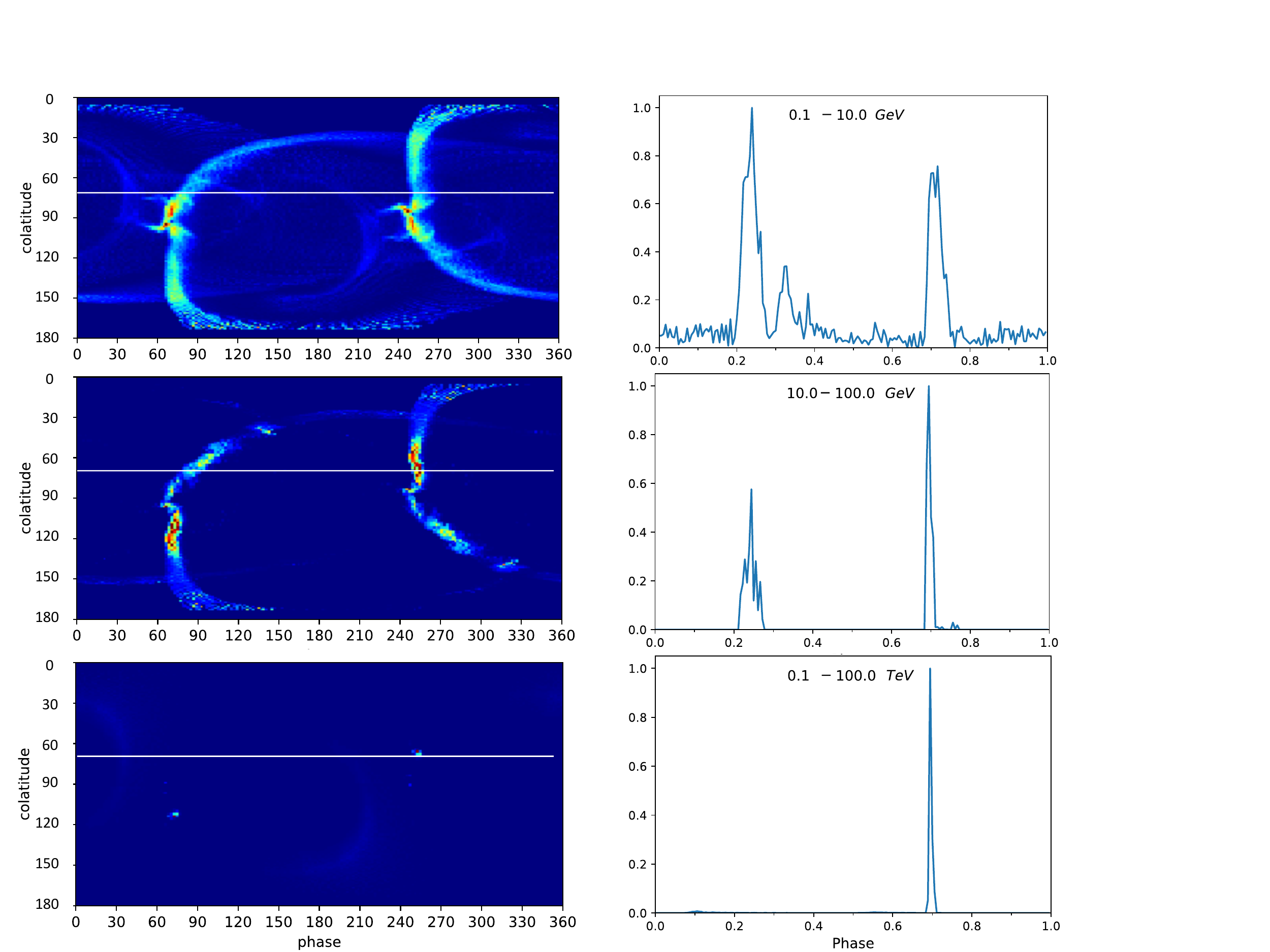}
\caption{Emission sky maps (intensity per solid angle for observer angle vs. rotation phase, both in degrees) and light curves for the Vela model shown in Figure~\ref{fig:Vela_spec1}, for three different energy ranges as labeled.  The light curves assume a viewing angle $\zeta = 70^\circ$, as indicated by the horizontal white lines in the sky maps.}
\label{fig:VelaLC}
\end{figure}

In Figure \ref{fig:VelaLC} we show the emission sky maps (intensity vs. $\zeta$ vs.\ $\phi$) in three different energy ranges, as indicated in the figure. Also shown are light curves for the same energy ranges, that are horizontal cuts through the sky maps at $\zeta = 70^\circ$.  The sky maps display bright caustics in the emission pattern that are characteristic of radiation from the FF current sheet (\citealt{Bai2010,Contopoulos2010}, Barnard et al., submitted).  The caustic patterns result from the particle acceleration being confined near the current sheet and their trajectories becoming nearly radial outside the light cylinder in the IOF.  The photons which they emit parallel to their motion then arrive to the observer at the same phase, producing a sharp pulse each time our line of sight crosses the current sheet.  Current sheet crossings can occur once, twice or not at all, depending on the viewing angle.  In these models (just like in Two-Pole-Caustic/SG models), light curves with two peaks are the result of emission from opposite rotational hemispheres, in contrast to traditional OG models \citep{Romani1995} where the emission from both peaks originates from the same hemisphere.   The particular pattern that appears in Figure \ref{fig:VelaLC} results from our assumption that $E_\parallel$ is constant in both radius and azimuth in the current sheet.  Somewhat different emission patterns can appear for cases where $E_\parallel$ varies with radius and azimuth in dissipative magnetosphere \citep{Kalapotharakos2014} or PIC \citep{Kalapotharakos2018} models, for example.  The light curves in those models provide a better match to the observed phases of the $\gamma$-ray peaks.

Even though we have assumed a uniform accelerating field in the current sheet, the intensity of emission varies along the caustics and the intensity pattern varies with energy.  In the lowest energy range, the emission is more uniform along the caustic than in the higher energy ranges where the highest-energy emission becomes more confined to particular locations.   These changing patterns of emission produce peaks  of similar flux at lower energy but increasingly unequal flux with increasing energy, with the first peak disappearing at the highest (TeV) energies (i.e., a continuation of the P1/P2 effect seen in the GeV band).  The caustics also narrow with increasing energy, producing a narrowing of the peaks.  Both of these effects, a harder second peak and a narrowing of both peaks with energy, are observed in many \textit{Fermi} pulsar light curves.  Similar results are presented by Barnard et al. (submitted). Emission primarily in the second peak at energies above 10 GeV is observed for Vela, as well as Geminga and B1706$-$44 (see \S \ref{sec:Geminga} and \ref{sec:B1706}).

Since we are also assuming that the MeV - GeV emission is SC, which is primarily CR at the energies displayed in Figure \ref{fig:VelaLC}, the variation in intensity is caused by variations in local radius of curvature, as found by Barnard et al. (submitted).  This is confirmed in Figure \ref{fig:VelaRho}, which shows a sky map of the maximum radius of curvature, $\rho_{\rm c}^{\rm max}$, of the trajectories of particles that emit photons in each direction $\zeta$ and $\phi$.  As in the intensity sky maps, particles on a range of different field lines and radii can emit photons in the same direction.  First of all, we notice that the sky pattern of $\rho_{\rm c}^{\rm max}$ closely reflects the caustic emission sky pattern.  This indicates that the largest curvature radii occur in and near the current sheet at radii outside the light cylinder where the main caustics form.  Also observe that the $\rho_{\rm c}^{\rm max}$ pattern is narrower than the emission caustics, indicating that the largest $\rho_{\rm c}$ values are more confined to the current sheet.  The bottom panel in Figure \ref{fig:VelaRho} shows a cut at $70^\circ$ through the sky map above to reveal that $\rho_{\rm c}^{\rm max}$ in the second light curve peak (P2) is about three times higher than that in the first peak (P1).  This figure then explains both the decreasing P1 to P2 ratio and the narrowing of the peak width with increasing energy as a characteristic of CR in a FF field geometry, where the spectral cutoff of CR is predicted to be
\be  \label{eq:Ec}
E_{\rm CR} \propto {E_\parallel}^{3/4}\,\rho_{\rm c}^{1/2}.
\ee
The above expression holds when the emitting particles are in radiation-reaction limit where energy losses balance acceleration gains.  
Figure 3 of HK15 shows that accelerating particles in the current sheet have reached the curvature radiation-reaction (CRR) regime.
The higher cutoff energy for P2 is due to the higher energy of the emitting particles in this regime,
\be  \label{eq:gammaCRR}
\gamma_{\rm CRR} = \left(\frac{3\,E_\parallel \,\rho_{\rm c}^2}{2e}\right)^{1/4},
\ee
given the larger value of $\rho_{\rm c}$ in this case.

Figure \ref{fig:Vela_PRspec} shows phase-resolved spectra for P1 and P2, illustrating that both the SC and ICS spectra of P2 are significantly harder.  It is apparent that the P2 spectrum has a cutoff at higher energy and this is reflected in the primary ICS spectrum.  Since the high-energy cutoff  of the P2 SC spectrum is due to the higher energy of the particles, these same particles are radiating an ICS spectrum that extends to higher energy.   

The decreasing P1/P2 ratio with increasing energy is dependent on $\alpha$ and $\zeta$.  For example, in Figure~\ref{fig:VelaRho} the value of $\rho_{\rm c}$ in P1 can be larger, smaller or similar to the $\rho_{\rm c}$ in P2, for different $\zeta$ values.  By symmetry, the ratio of $\rho_{\rm c}$ for P1 and P2 approaches 1 for $\zeta$ closer to $90^\circ$.  For smaller $\alpha$, the energy dependence of P1/P2 will be smaller for a wider range of $\zeta$ (see \S \ref{sec:Crab}).

\subsection{Crab pulsar} \label{sec:Crab}

The model SED for the Crab pulsar is shown in Figure~\ref{fig:Crab_spec} for the pair spectrum with a power-law extension (solid line in Figure \ref{fig:pair}) and two different viewing angles. The pair SR component matches the optical through hard X-ray data reasonably well and requires a large pair multiplicity that is both suggested by nebular models and is theoretically possible in PC cascade models \citep{Timokhin2015}. We find that the pair energy spectrum with a power-law extension, combined with the constant pair pitch angle, can account for the COMPTEL soft $\gamma$-ray points plus the lower part of the {\it Fermi} spectrum and produces a pair SSC spectrum that matches the MAGIC VHE points.  The primary SC spectrum, combined with the high-energy part of the pair SR spectrum, can match the {\it Fermi} points up to about 50 GeV.  The flux of the pair SSC component exceeds that of the primary SC component above 50 GeV. In this model, the MAGIC spectrum is not an extension of the primary SC spectrum, predicting a possible dip or discontinuity between the \textit{Fermi} and MAGIC spectra.  Our model also predicts a primary $\gamma$-$\gamma$ attenuated ICS component peaking around $10 - 20$ TeV that is potentially detectable by HAWC.  However, HAWC is primarily designed to operate in imaging mode and has presently not implemented searches for pulsed emission (B. Dingus, priv. comm.).  There are several MAGIC points and upper limits at the level our primary ICS spectral prediction, implying that more observation time by MAGIC may be able to detect part of this component.  The pair SR spectrum for $\zeta = 60^\circ$ matches the optical and X-ray data slightly better than the $\zeta = 72^\circ$ pair SR spectrum, but the $\zeta = 72^\circ$ pair SSC spectrum better matches the MAGIC points.  

Figure \ref{fig:CrabLC} shows the emission sky maps and light curves for $\zeta = 72^\circ$ and for different energy ranges, one in the optical range and three at $\gamma$-ray energies.  The model optical light curve shows two narrow peaks that are in phase with those of the $\gamma$-ray light curves, a result of locating all the emission in the high-altitude separatrix and current sheet.  Even though the pairs and primaries are injected on different but adjacent sets of magnetic field lines at the NS surface, the radiation they emit near the current sheet produces light curve peaks at the same phase.  This is because the trajectories of particles outside the light cylinder become nearly radial. Radiation from these same sets of field lines inside the light cylinder would produce light curves at slightly different phases.  As was also seen for our Vela model, the emission intensity distribution along the caustics changes with energy.  However, for this value of $\alpha = 45^\circ$ and $\zeta = 72^\circ$ the relative intensity variation of P1/P2 is weaker.  The decrease in P1/P2 flux with energy up to 50 GeV is slower than for Vela although we see that P1/P2 is smaller in the $10 - 50$ GeV light curve.  Indeed, the \textit{Fermi} Crab light curves show a slower energy evolution compared to that of Vela.

\subsection{Geminga pulsar} \label{sec:Geminga}

In Figure~\ref{fig:Gem_spec} we show model SEDs for the Geminga pulsar for $\alpha = 75^\circ$, $\zeta = 55^\circ$ and two different variations of $R_{\rm acc}$.  The solid curve assumes two values inside and outside the light cylinder,  $R_{\rm acc}^{low} = 0.04$ and $R_{\rm acc}^{high} = 0.15$, while the dashed spectra assumes a single value $R_{\rm acc}^{low} = R_{\rm acc}^{high} = 0.15$ at all radii.  
Our assumed radio flux of 1000 mJy can account for the observed optical/UV emission through the pair SR component.  The pair SR is very sensitive to $\zeta$ for $\zeta \lsim 50^\circ$ since our line-of-sight must cut through enough of the low-altitude emission from the pairs which is within about $30^\circ$ of the magnetic pole at $75^\circ$.  A viewing angle of $\zeta \sim 50^\circ - 55^\circ$ produces a flux of pair SR and shape of primary SC emission that best matches the data.  The two $E_\parallel$ models produce similar primary SC spectra above about 1 GeV, where the CR regime dominates, but differ markedly below 1 GeV where SC becomes a mix of SR and CR.  The model with the lower $E_\parallel$  below the light cylinder suppresses acceleration at lower altitudes so that  the primaries can temporarily maintain larger pitch angles and radiate a larger amount of SR.  Since the hard X-ray spectrum of Geminga is not well measured, the $E_\parallel^{low}$ value is presently unconstrained.
We see that the MAGIC spectrum that is an extension of the \textit{Fermi} spectrum is well explained as primary SC emission with no need to invoke any ICS component.  Indeed, our predicted ICS emission, both from pairs (whose flux is below the level of the plot) and primaries, is well below any present detection thresholds.  However, the two different models for $E_\parallel$ predict quite different primary ICS spectra, with the multi-valued $E_\parallel$ model ICS spectrum extending to lower energies and having a slightly larger flux.  Such an effect was also apparent in the Vela models presented by H18, where the observed hard-X-ray flux constrains the $R_{\rm acc}^{low} = 0.04$.

As is apparent from Figure \ref{fig:pair}, the computed Geminga pair spectrum is much lower than that of Vela, in total multiplicity and also in both low and high energy extent.  Thus the predicted pair SR spectrum for Geminga has a lower flux and SED energy peak, both by about two orders of magnitude, and a narrower energy range compared with Vela.  The light cylinder distance is also three times larger than for Vela, so that the the high-energy particles in the current sheet are farther from the soft photon source, making their local density lower.  So even though the primary SC flux is comparable to Vela, the lower soft-photon density produces a lower primary ICS component. The Geminga spectrum at soft X-ray energies is also dominated by a strong thermal component which is not shown.

Figure \ref{fig:GemLC} shows the sky maps and light curves in three different energy ranges, the lowest one in the \textit{Fermi} range, the middle one for the MAGIC range and the third for TeV energies.  As in the case of Vela, for which we assumed the same $\alpha$ value, there is a strong dependence of the sky map intensity distribution on energy.  This produces a strong decrease in the P1/P2 ratio with increasing energy, such that P1 should disappear above $\sim 20$ GeV.  Indeed, this is the case for both \textit{Fermi} and MAGIC light curves where P1 is not detected above 10 GeV, although the observed narrowing of the peaks is not as pronounced as indicated by our prediction.  

\subsection{PSR B1706$-$44}  \label{sec:B1706}

Figure \ref{fig:B1706_spec} shows two SED models for PSR B1706$-$44, a Vela-like pulsar, one with $\alpha = 45^\circ$ and $\zeta = 53^\circ$ chosen to match the \textit{Fermi} light curve and another with $\alpha = 30^\circ$ and $\zeta = 60^\circ$ that better matches the {\it Chandra} spectrum. The {\sl Fermi} light curve is very different from that of Vela, with a narrow spacing of the observed $\gamma$-ray peaks that is better modeled by smaller inclination angles where our line-of-sight can cut through the current sheet caustics at a grazing angle as shown in Figure \ref{fig:B1706LC}.  There is no pulsed optical emission detected for this pulsar, since its distance is much larger than Vela's or Geminga's but there is a soft power-law X-ray detection from \citet{Gotthelf2002} which we use to match the flux of the pair SR component.  The predicted pair SR spectrum has a higher flux (by two orders of magnitude) and SED peak energy (by a factor of 10 to 100) than for Geminga, primarily because its pair spectrum is higher in multiplicity and extends to higher energy.  Although the B1706$-$44 pair spectrum is very similar to that of Vela, the pair SR spectral flux is somewhat lower since the distance is larger and the radio luminosity is lower.  However, the intrinsic luminosity of both the pair SR and primary SC must be larger than Vela's given that the \textit{Fermi} peak SED flux is less than a factor of ten lower.  Therefore, the  luminosity of the primary ICS is predicted to be higher than Vela's and it is only its larger distance and relatively lower predicted optical/IR  flux that leads to the flux level being below detection thresholds.  However, we predict that the flux of the primary ICS component is a factor of ten above that for Geminga. The two different model geometries produce two very different primary SC spectra since changing $\alpha$ and $\zeta$ changes the particle trajectory radii of curvature which changes the dynamics and spectrum of the SC radiation.  In some cases, the spectrum is more SR dominated at low energies while in others it is more CR dominated.  Therefore a change in geometry, as well as a difference in $E_\parallel$ distributions as in two Geminga models we presented, can produce different primary SC spectra.
We note that the primary ICS spectrum for the $\alpha = 30^\circ$ model (dashed line) reaches a bit higher in flux, reflecting the low-energy enhancements of the primary SC spectrum and particle energy distribution. H.E.S.S.-II has detected pulsed emission up to 70 GeV which we can explain as primary SC emission, as for Geminga.  The pair SSC component is very weak and is not predicted to produce an additional component at VHE energies.  

We show the energy-dependent sky maps and light curves for B1706$-$44 in Figure~\ref{fig:B1706LC}.  The first two energy ranges are chosen to match the \textit{Fermi} and H.E.S.S.-II detection ranges.  As is the case for the other sources, the caustics in the emission sky map narrow with increasing energy and for this model $\zeta$, the P1/P2 ratio decreases with increasing energy.  This behavior is in agreement with the H.E.S.S.-II 10 - 70 GeV light curves in which P2 dominates.  However, the narrowing of the peaks is not so apparent, either in the \textit{Fermi} \citep{Abdo2013} or H.E.S.S.-II light curves \citep{SpirJacob2019}.  In the highest energy band, $> 70$ GeV, we predict that only P2 will be present.

\subsection{PSR J0218+4232} \label{sec:J0218}

Our final source model is for J0218+4232, an energetic MSP with a very hard X-ray spectrum that extends to at least the highest \textit{NuSTAR} energy 80 keV.  Figure~\ref{fig:J0218_spec} shows the model SED for inclination angle $\alpha = 60^\circ$ and viewing angle $\zeta = 65^\circ$, using the pair spectrum shown in Figure \ref{fig:pair}.  The pair SR is very different from that of the younger pulsars, reflecting the much more energetic pair spectrum.  The  SR SED peak is predicted to be near 10 MeV, which matches the hardness of the observed X-ray spectrum.  As a result, the optical/IR flux is predicted to be much lower than that of the younger pulsars by at least several orders of magnitude.  Consequentially, the pair SSC component peaks at a VHE energy around 50 GeV but at a flux level well below any current detection threshold.  The primary SC component can account for the higher-energy \textit{Fermi} points, but falls short of the lower energy ones. It is possible that a stronger SR regime of primary SC radiation, or a pair spectrum extending to higher energy producing a pair SR component extending to around 1 GeV could reach the \textit{Fermi} spectrum.  A telescope with sensitivity at 1 - 100 MeV energies could address this question.

Our model does not predict a detectable primary ICS component.  The SED peaks around 1 TeV, a decade lower than for the younger pulsars, due to suppression by Klein-Nishina effects, as none of the ICS occurs in the Thompson limit and the flux is also suppressed.  This effect was seen by HK15 who modeled pair SSC for two other energetic MSPs, B1821$-$24 and B1937+21, whose pair spectra are very similar to that of J0218+4232.  Our model therefore suggests that MSPs will not be good targets for VHE telescopes.  As shown in the figure, MAGIC has data totaling 90 hrs observing J0218+4232 at 20 GeV - 20 TeV, with only upper limits, although \textit{Fermi} detects pulsed emission up to 25 GeV \citep{Acciari2020}.  The discrepancy around 20 - 25 GeV is due to the high cosmic-ray background subtraction affecting MAGIC and other ground-based air-Cherenkov detectors.  

\section{Discussion}\label{sec:disc}

We have presented models of broadband emission, covering IR to VHE energies, from a number of rotation-powered pulsars assuming the global magnetic field structure of an FF magnetosphere.  While much of the method of calculation is very similar to that of HK15, the expanded photon energy range that is modeled here (from 12.5 to 18 energy decades) fundamentally changes the results to reveal a number of new features.  Prime among these is the appearance of a VHE emission component at photon energy of 0.1 - 30 TeV from accelerated primary particles in the current sheet scattering very low energy photons produced by pair SR.  There was a hint of this component in Figures 5 and~6 of HK15, but it was cut off above the maximum modeled energy of 5 TeV and suppressed in flux by a cutoff in the pair SR spectrum below the minimum modeled energy 1 eV, preventing the primary particles from scattering IR photons in the Thompson limit.  Such a limitation was remedied for the Vela model in H18 and is now also for the Crab and several over sources in this paper.  
Another major improvement made possible by the expanded photon energy range is the ability to use the observed optical/UV spectrum to constrain the emission altitude and the multiplicity of the electron-positron pair SR component.  This constrains the IR to UV photon density at the source that then determines the primary ICS flux above 10 TeV, as well as the pair SSC component. Since the pairs are injected from the surface, there is no free parameter for the IR/UV emission; only the radio emission height is free.

\begin{figure} 
\includegraphics[width=200mm]{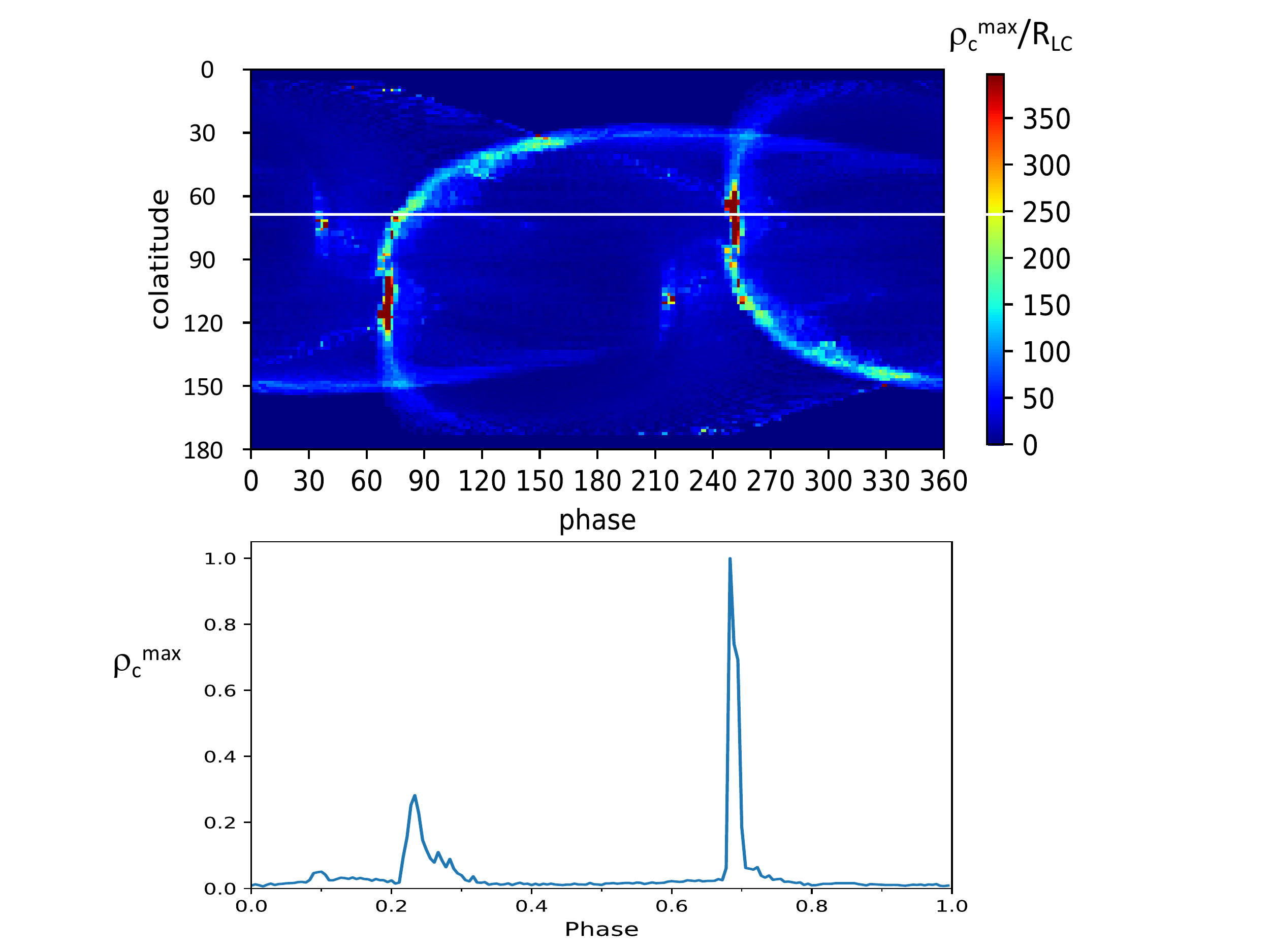}
\caption{Top: Sky map of particle trajectory maximum radius of curvature, $\rho_{\rm c}^{\rm max}$, in units of light cylinder radius, $R_{\rm LC}$.  Bottom: Slice through the sky map (solid white line) showing the variation of $\rho_{\rm c}^{\rm max}$ with pulse phase at $\zeta = 70^\circ$ (normalized to 1.0 at maximum).}
\label{fig:VelaRho}
\end{figure}

The results of our modeling predict that three main components contribute to pulsar VHE emission.  First, we have shown that emission up to 100 GeV can be produced by {\bf primary SC} and the tail of this spectrum can account for the pulsed emission observed by H.E.S.S.-II from Vela and B1706$-$44, by MAGIC from Geminga and by \textit{Fermi} for J0218+4232.   The extended sub-exponential, power-law-like high-energy tail of this component results from emission of particles in the CR regime of SC with different radiation-reaction limited energies.  The particle trajectory radii of curvature vary at different altitudes along the current sheet, producing a range of CR cutoffs (see Eqn [\ref{eq:Ec}]) that all combine in the caustic peaks.  The emission above 10 GeV is observed most strongly in the second light curve peak, extending the P1/P2 ratio decrease with increasing energy observed in many pulsars at lower energies by \textit{Fermi}.  This is caused by the different maximum radii of curvature of the particle trajectories that produce the two peaks (Figure \ref{fig:VelaRho}).
ICS may also account for this emission but its spectrum would need to join smoothly to the \textit{Fermi}  spectrum and the P1/P2 trend might not be smooth or may occur at higher energy.

The second predicted VHE emission component comes from {\bf pair SSC}.  The SSC from pairs produces a broad SED whose peak typically occurs between $1 - 10$ GeV for young pulsars and around $100$~GeV for MSPs.  In all of the source models we have presented in this paper, the pair SSC component flux lies well below, and is thus obscured by, the primary SC emission below 100 GeV.  Above 100 GeV, the pair SSC emission extends to around 1 TeV and may be detectable in some pulsars like the Crab and Crab-like pulsars.  A pair SSC component high enough to be detectable requires 1) high pair multiplicity (the SSC flux is proportional to the square of $M_+$) 2) high magnetic field strength at the light cylinder, $B_{\rm LC}$ (to boost the SR level), and 3) low pair energies so that the scattering is not in the extreme Klein-Nishina regime.  

\begin{figure} 
\includegraphics[width=180mm]{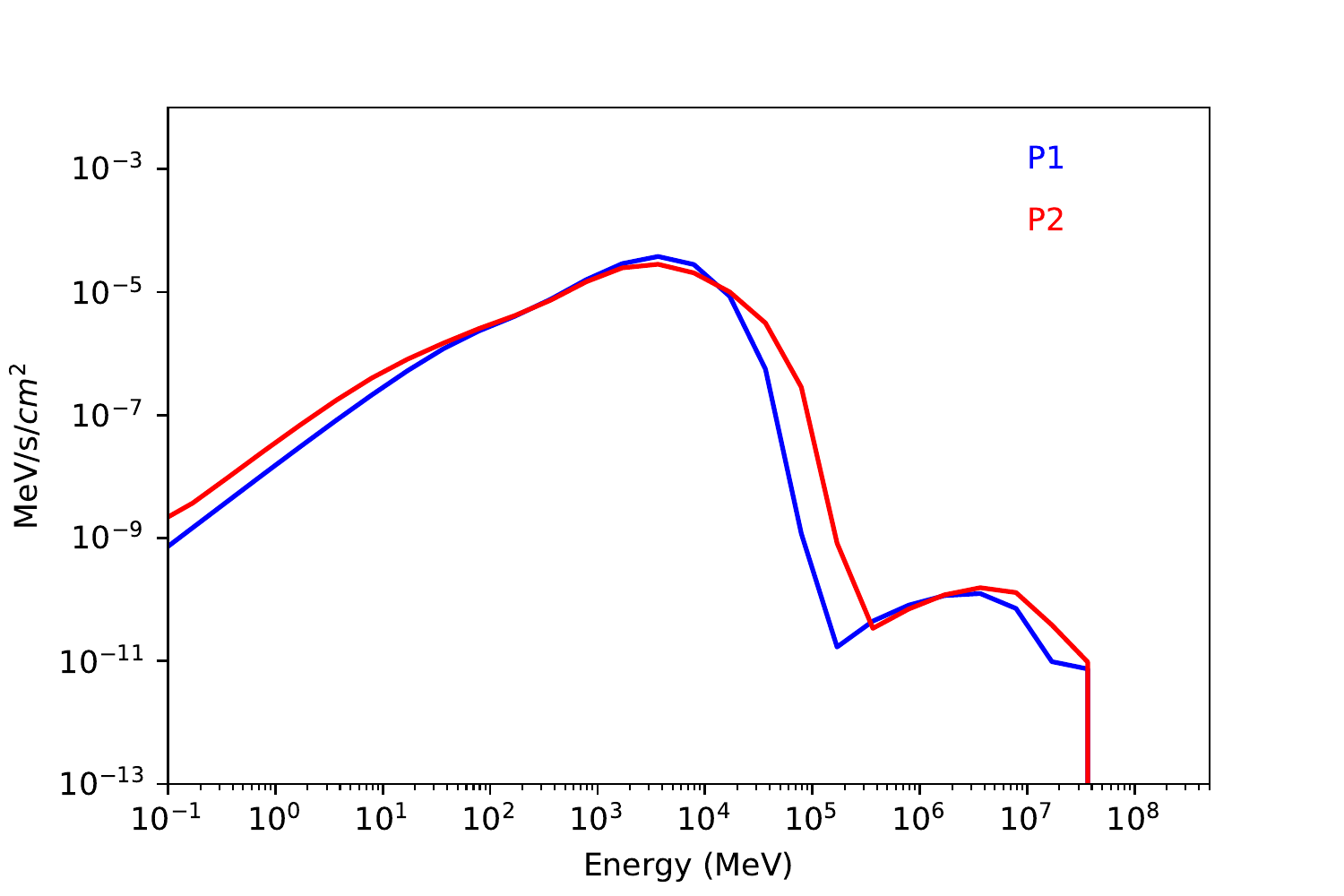}
\caption{Model SED for the Vela model of Figure~\ref{fig:Vela_spec1} for the phase ranges of the first and second peaks, P1 and P2. One can clearly see the relatively larger SC and ICS spectral cutoffs for P2.}  
\label{fig:Vela_PRspec}
\end{figure}

\begin{figure} 
\includegraphics[width=180mm]{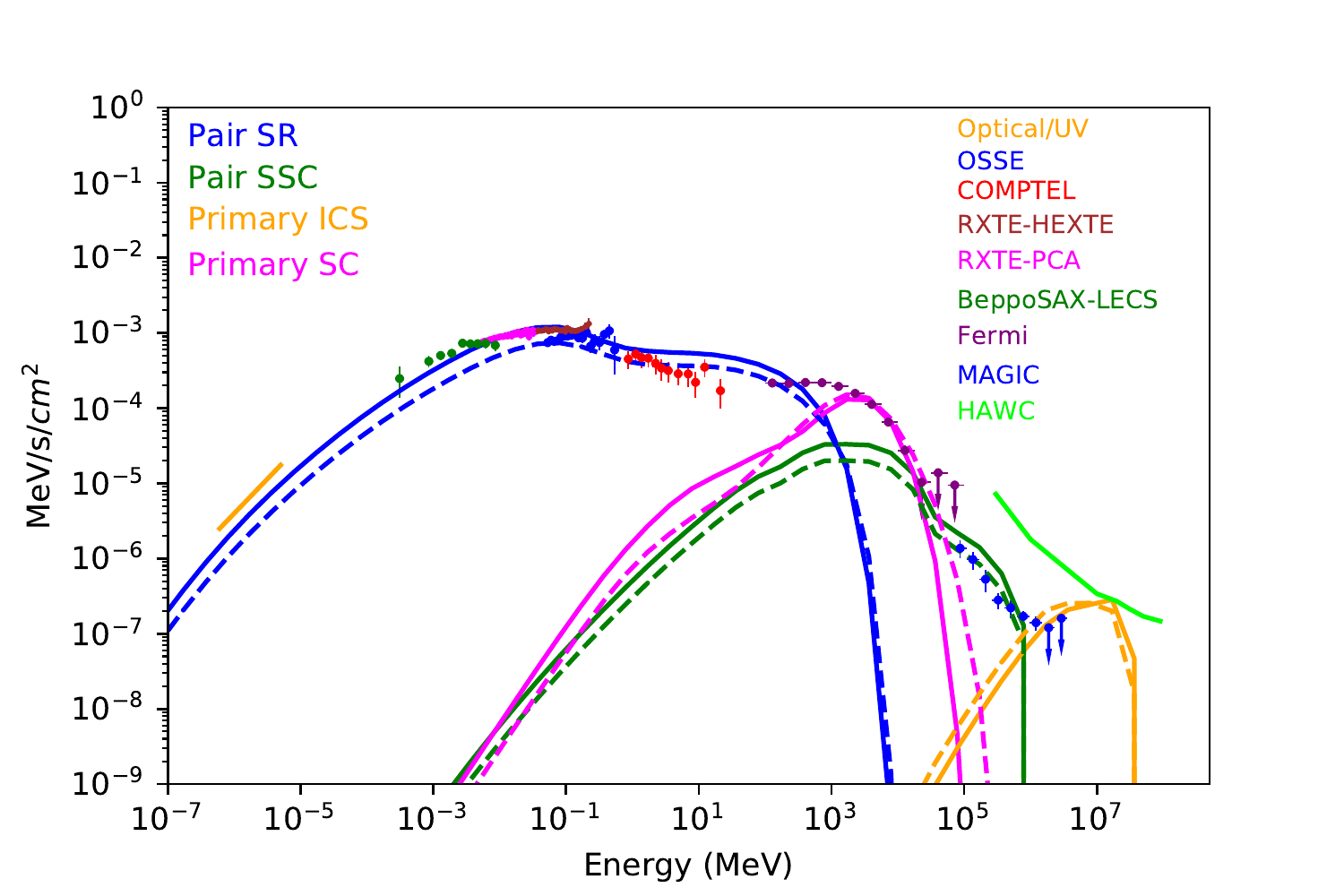}
\caption{Model SED for the Crab pulsar for inclination angle $\alpha = 45^\circ$ and two different viewing angles, $\zeta =  60^\circ$ (solid lines) and $\zeta =  72^\circ$ (dashed lines) for the pair spectra with power-law extension shown in Figure~\ref{fig:pair},  assuming $M_+ = 3 \times 10^5$.  Data points are from \citet{Kuiper2001}, \citet{Abdo2013} (http://fermi.gsfc.nasa.gov/ssc/data/access/lat/2nd\_PSR\_catalog/), \citet{Ansoldi2016} and \citet{Sollerman2019}.  The HAWC sensitivity curve for 50 hr of observation is also shown \citep{Abeysekara2017}}.  
\label{fig:Crab_spec}
\end{figure}

\begin{figure} 
\includegraphics[width=200mm]{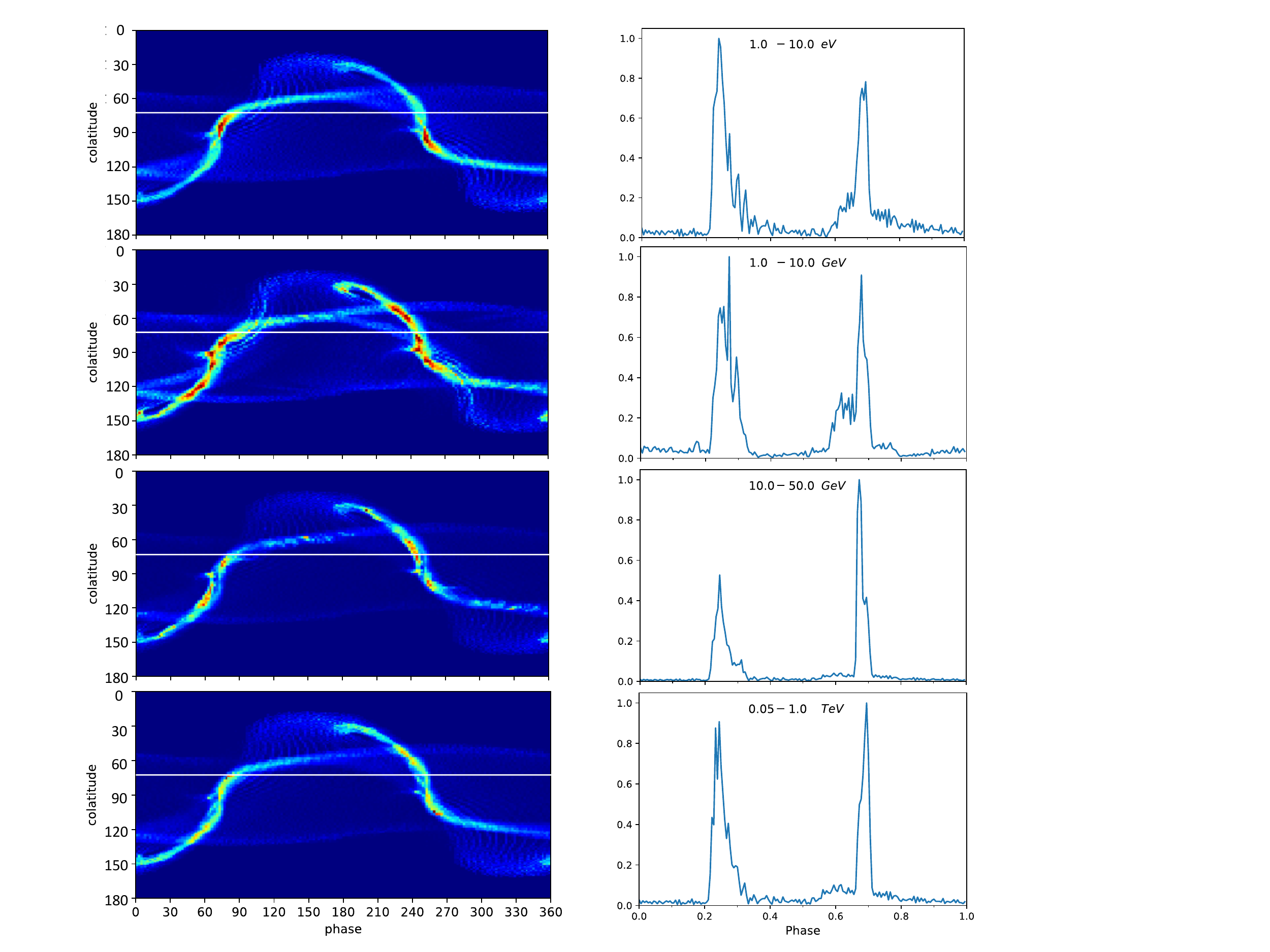}
\caption{Emission sky maps and light curves for the Crab extended power-law pair model shown in Figure \ref{fig:Crab_spec}, for four different energy ranges as labeled.  The light curves assume a viewing angle $\zeta = 72^\circ$, as indicated by the horizontal white lines in the sky maps.}
\label{fig:CrabLC}
\end{figure}

\begin{figure} 
\includegraphics[width=180mm]{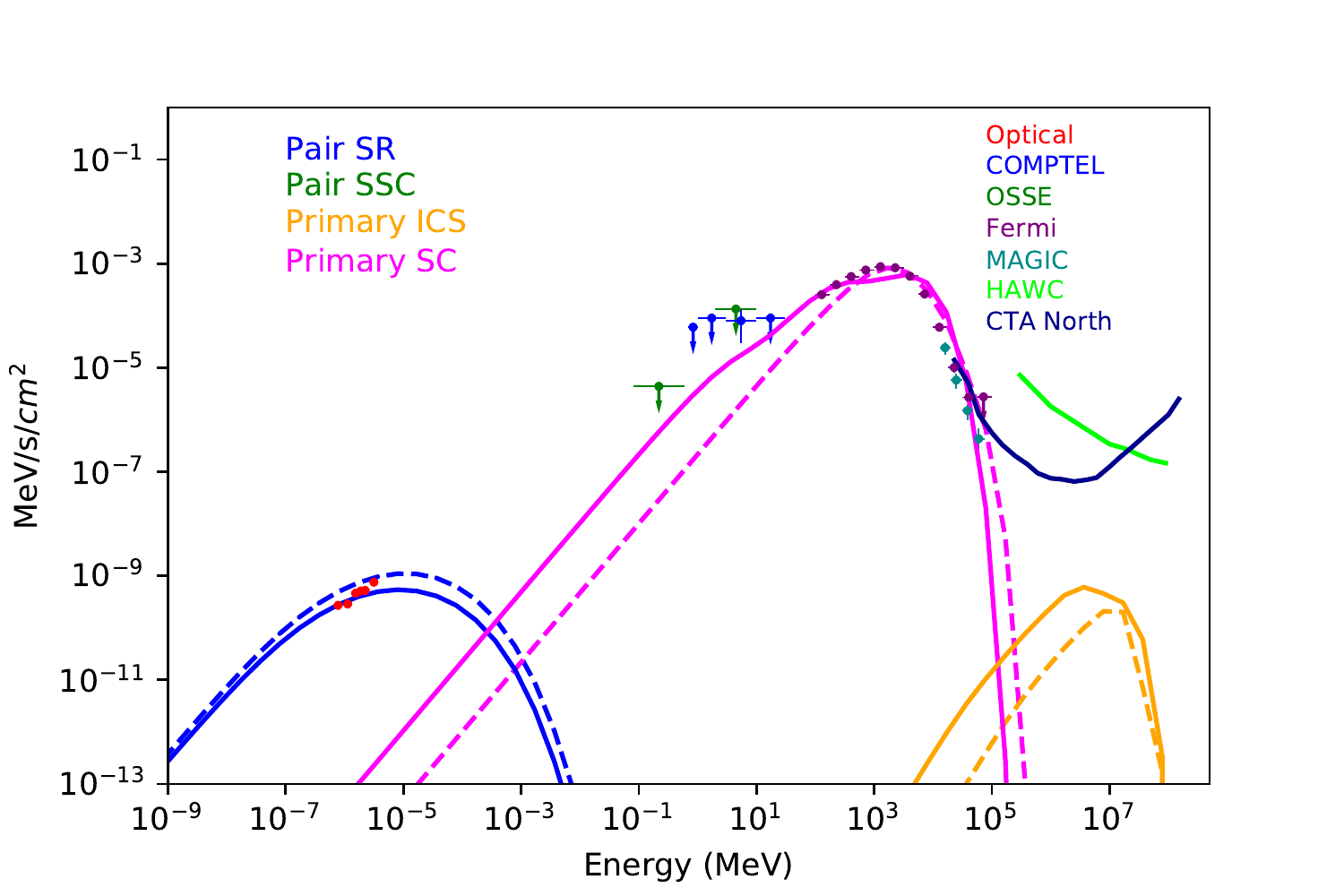}
\caption{Model SED for the Geminga pulsar for inclination angle $\alpha = 75^\circ$, viewing angle $\zeta = 55^\circ$, using the solid pair spectrum (dipole field) shown in Figure~\ref{fig:pair} with $M_+ = 2 \times 10^4$.  The solid line spectra are for $R_{\rm acc}^{low} = 0.04$ and $R_{\rm acc}^{high} = 0.15$, while the dashed spectra are for $R_{\rm acc}^{low} = R_{\rm acc}^{high} = 0.15$.  Data are from \citet{Kargaltsev2005}, \citet{Shibanov2006}, \citet{Kuiper1996}, \citet{Abdo2013} (http://fermi.gsfc.nasa.gov/ssc/data/access/lat/2nd\_PSR\_catalog/), 
and \citet{Acciari2020}.  HAWC \citep{Abeysekara2017} and CTA North \citep{Acharya2019} sensitivity curves for 50 observation hr are also shown.}
\label{fig:Gem_spec}
\end{figure}

\begin{figure} 
\includegraphics[width=200mm]{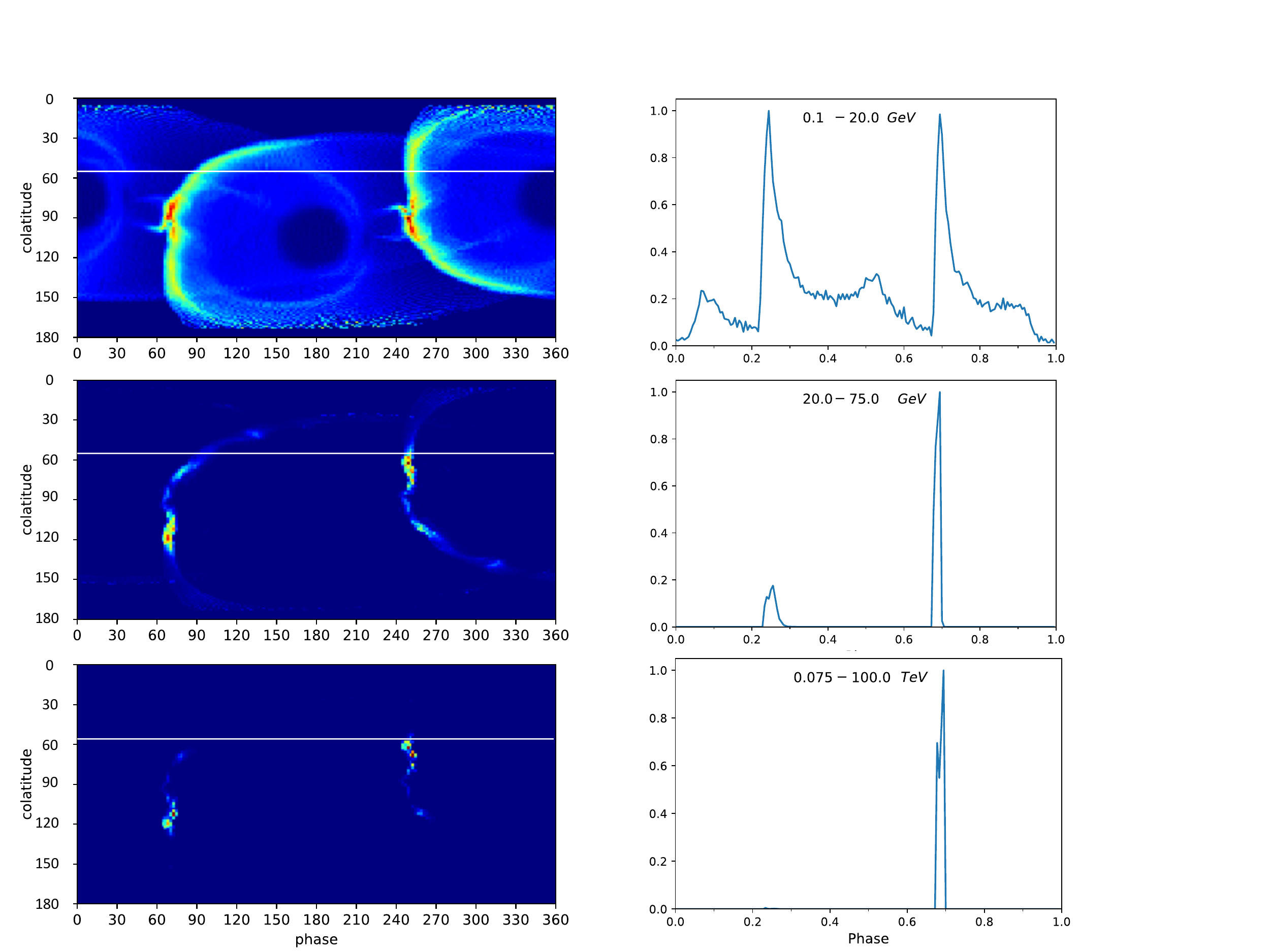}
\caption{Emission sky maps and light curves for the Geminga model shown in Figure \ref{fig:Gem_spec}, for three different energy ranges as labeled.  The light curves assume a viewing angle $\zeta = 55^\circ$, as indicated by the horizontal white lines in the sky maps.}
\label{fig:GemLC}
\end{figure}

\begin{figure} 
\includegraphics[width=180mm]{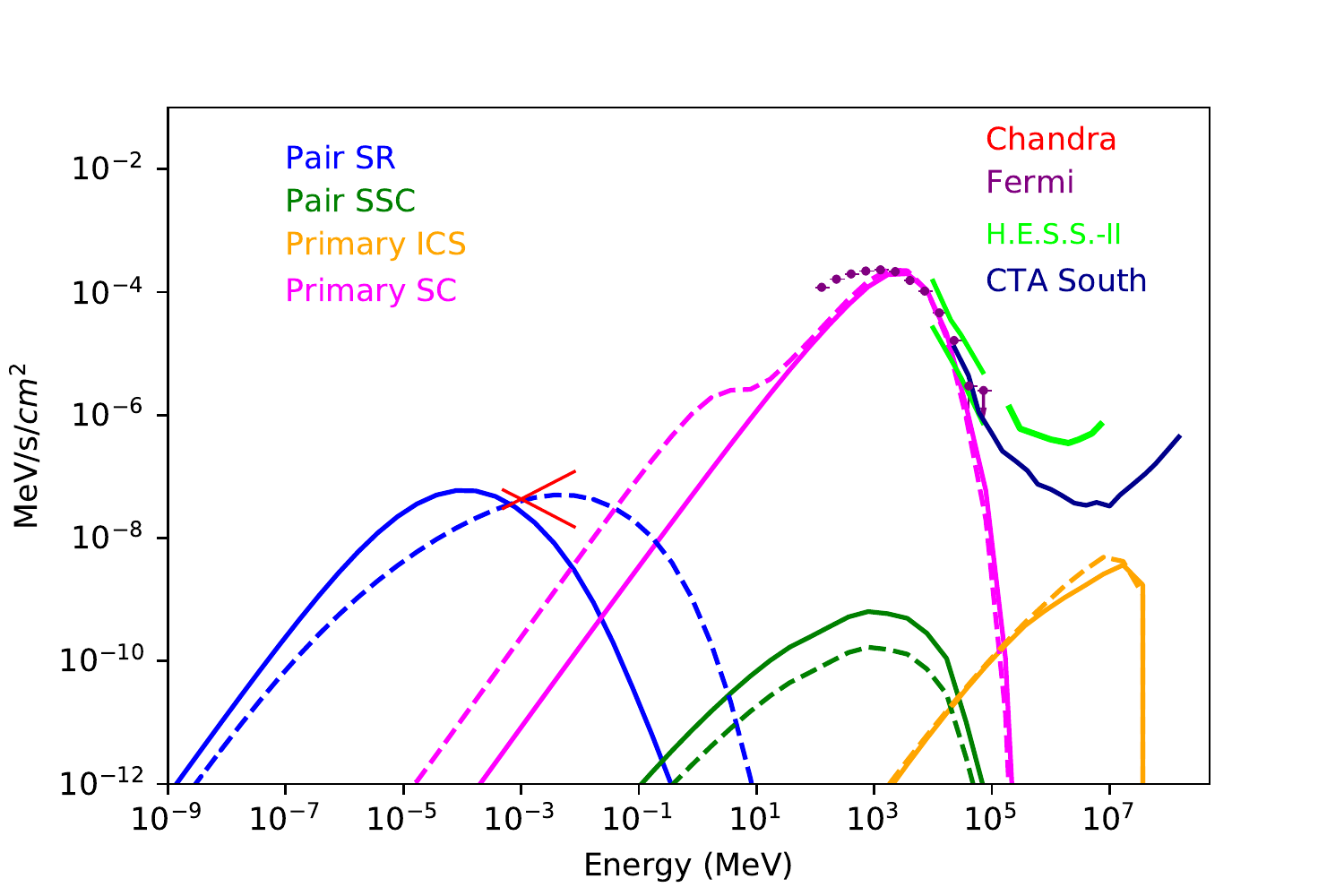}
\caption{Model SED for PSR B1706$-$44 for inclination angle $\alpha = 45^\circ$ and viewing angle $\zeta =  53^\circ$ (solid lines) and $\alpha = 30^\circ$ and $\zeta = 60^\circ$ (dashed lines), using the solid pair spectrum (dipole field) shown in Figure~\ref{fig:pair}, with $M_+ = 6 \times 10^4$.  Data are from \citet{Gotthelf2002}, \citet{Abdo2013} (http://fermi.gsfc.nasa.gov/ssc/data/access/lat/2nd\_PSR\_catalog/) and \citet{SpirJacob2019}.  H.E.S.S.-II \citep{Holler2015} and CTA South \citep{Acharya2019} for 50 observation hr are also shown.}  
\label{fig:B1706_spec}
\end{figure}

\begin{figure} 
\includegraphics[width=200mm]{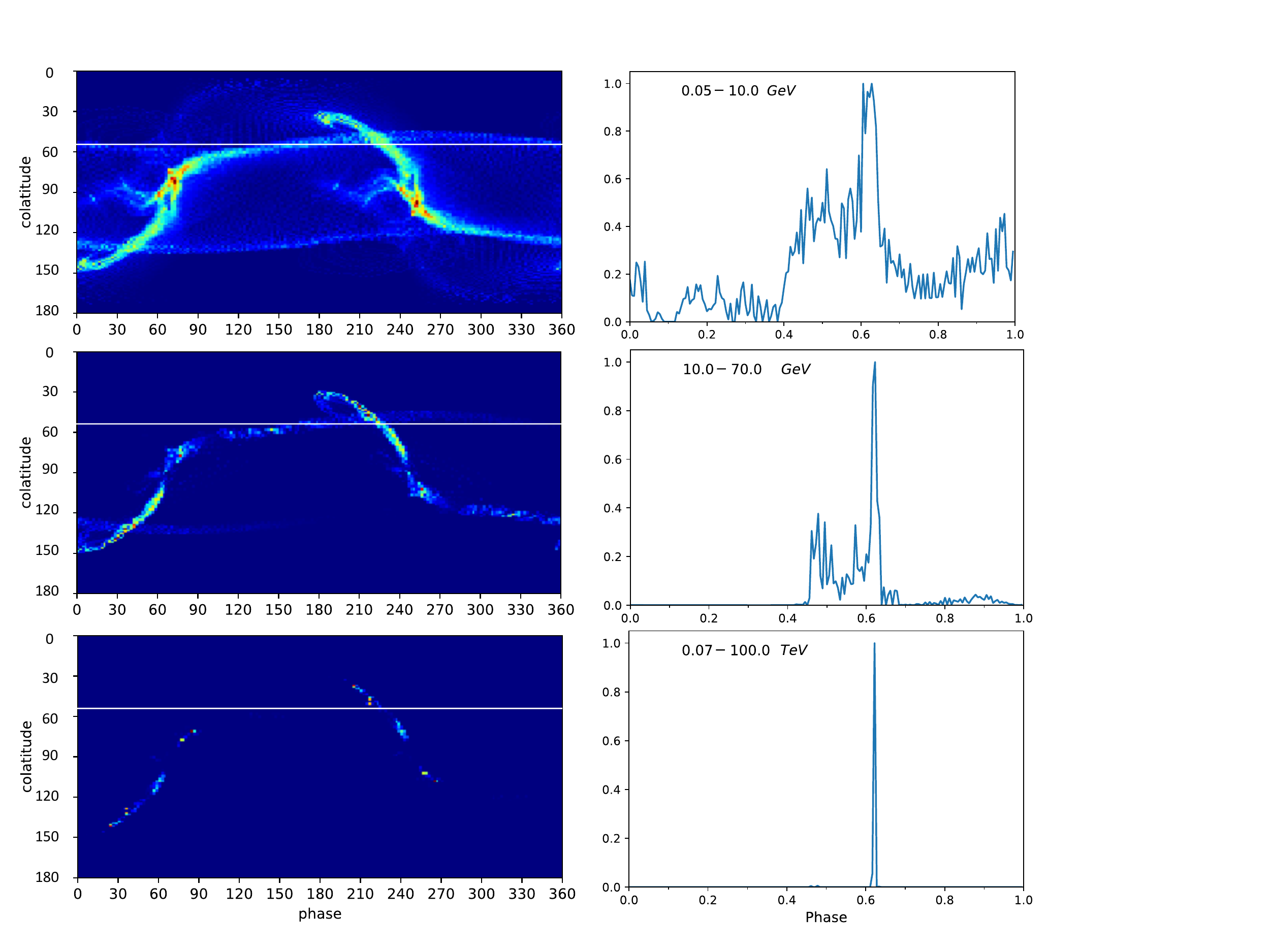}
\caption{Emission sky maps and light curves for the B1706-44 model shown in Figure \ref{fig:B1706_spec}, for three different energy ranges as labeled.  The light curves assume inclination angle $\alpha = 45^\circ$ and viewing angle $\zeta = 53^\circ$, as indicated by the horizontal white lines in the sky maps.}
\label{fig:B1706LC}
\end{figure}

\begin{figure} 
\includegraphics[width=180mm]{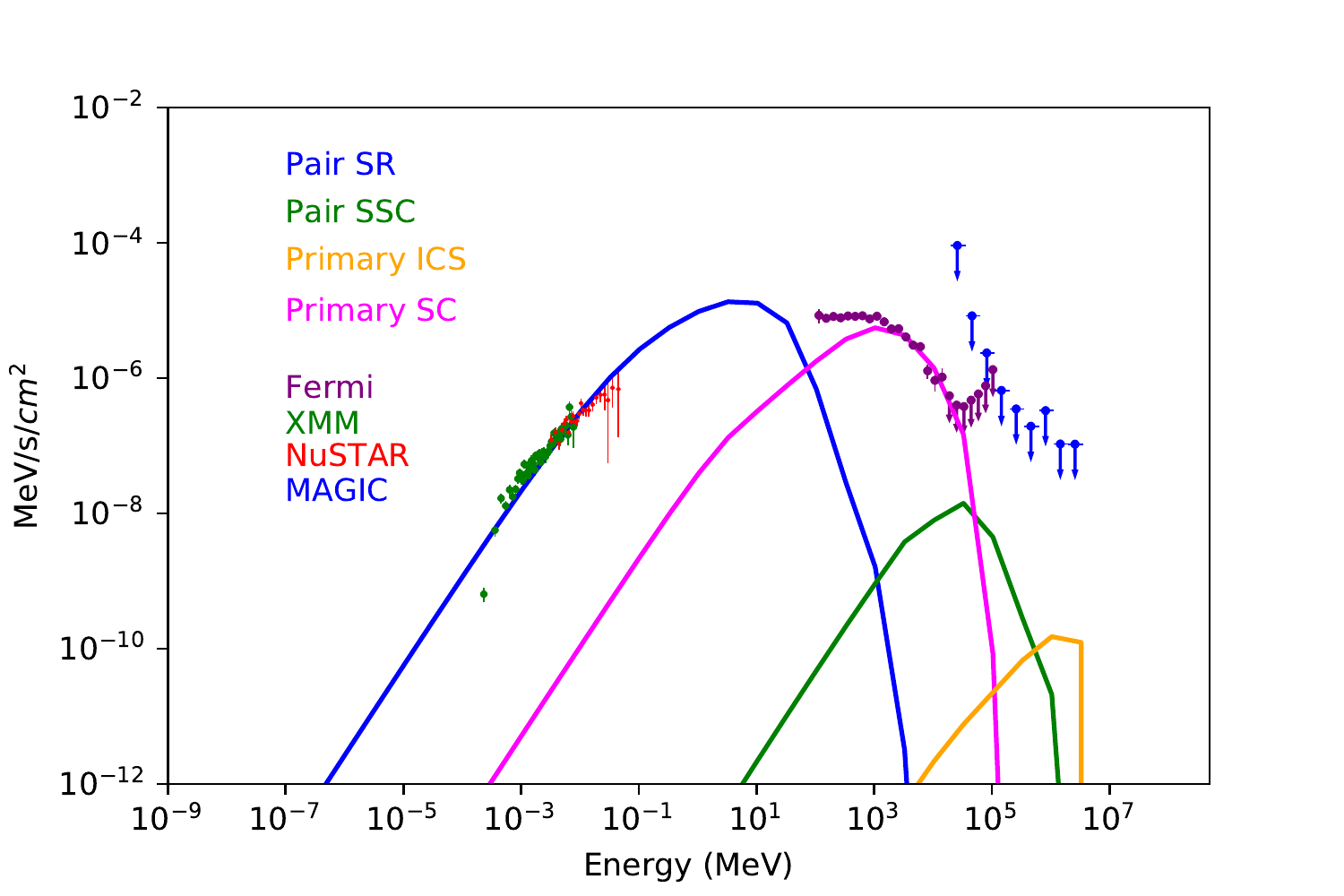}
\caption{Model SED for PSR J0218+4232 for inclination angle $\alpha = 60^\circ$ and viewing angle $\zeta =  65^\circ$, using the solid pair spectrum (dipole field) shown in Figure \ref{fig:pair} with $M_+ = 3 \times 10^5$.  \textit{XMM}-Newton and NuSTAR data is from \citet{Gotthelf2017} and \textit{Fermi} and MAGIC data is from \citet{Acciari2020}.}  
\label{fig:J0218_spec}
\end{figure}

The third predicted VHE emission component comes from {\bf primary ICS}, radiation from the most highly accelerated particles in the current sheet scattering low--energy pair SR.  Our models predict primary ICS components for all pulsars but for many the flux is well below current detection thresholds.  This component depends on all five free parameters of our model and also requires treating a very large range of photon energies, at least 18 decades.  For some pulsars with large IR fluxes, it may be necessary to include one or two decades below what we have considered here.  The most important constraint that can be derived from a measurement of this component is the maximum energy of the accelerated particles, which by conservation of energy is simply equal to the high-energy cutoff of the SED.  Even without a precise measurement of the SED high-energy cutoff, the maximum photon energy detected puts a lower limit on the maximum particle energy.  Particles with energies above 10 TeV are radiating in the CR regime of SC, thus ruling out models suggesting that the \textit{Fermi} GeV emission is SR \citep{Cerutti2016,Philippov2018}.  
Most of the primary ICS emission should appear in the second light curve peak for most values of  $\alpha$ and $\zeta$.  Since it has been shown that P2 is radiated by the highest-energy particles (Barnard et al., submitted), those same particles will produce the highest ICS energies, as shown in Figure \ref{fig:Vela_PRspec}.
Pulsar characteristics needed for detectable primary ICS emission are 1) particle energies at least 10 TeV 2) high IR/optical flux (this is not necessarily correlated with pair multiplicity but with high radio luminosity) and 3) a relatively small distance between the low-altitude pair emission (radio altitude) and the current sheet, favoring shorter pulsar periods.

In our model, the primary ICS spectrum is predicted to sharply cut off at the maximum particle energy.  The cutoff energy is constrained by fitting the {\sl Fermi} high-energy cutoff, which is the cutoff in the CR spectrum of the same particles.  Since the FF magnetic field structure determines the curvature radii of the particle trajectories, the {\sl Fermi} spectral cutoff sets the maximum particle energy assuming the particles have reached radiation-reaction limit.  Therefore, with the assumptions of a FF field, particles in CR-reaction limit and emission from the current sheet, the primary ICS cutoff is constrained to be 20-30 TeV.

In addition to a constraint on maximum particle energy, a detection of the primary ICS component combined with the cutoff in the SC/CR spectrum would provide separate constraints on maximum radius of curvature and $E_\parallel^{high}$.  The primary ICS and SC cutoffs constrain different combinations of $\rho_{\rm c}^{\rm max}$ and $E_\parallel^{high}$, the primary ICS cutoff through $\gamma_{\rm CRR}$ by Eqn (\ref{eq:gammaCRR}) and the primary SC cutoff through $E_{\rm c}$ through Eqn (\ref{eq:Ec}).  Measurement of the primary ICS cutoff for a variety of pulsars with different \textit{Fermi} light curves could test whether the P1/P2 ratio decrease with energy is caused by spatial variation of $\rho_{\rm c}^{\rm max}$ or  $E_\parallel^{high}$, or a combination of both, in addition to probing the effect of $\alpha$ and $\zeta$ on this trend.  Finally, the shape of the primary ICS spectrum is dependent on primary particle energies as a function of altitude, since it is reflected in the shape of the SC spectrum.   Measurement of the ICS spectrum below the cutoff can therefore constrain the primary particle energy distribution.

Given the model requirements for production of the different VHE emission components, we can make predictions for the pulsars that offer the best prospects for detection by VHE telescopes.  For the pulsars in the Second \textit{Fermi} Pulsar Catalog \citep{Abdo2013}, there is a very weak correlation between the GeV cutoff $E_{\rm c}$  (with a range less than a decade) and magnetic field at the light cylinder (with a range of about 3 decades).  The sub-100 GeV primary SC component is thus not likely to be very sensitive to parameters of individual pulsars, so the VHE detectability will scale with the \textit{Fermi} flux (as this indicates a higher flux of primary particles).  Consequently, several of the pulsars with the largest \textit{Fermi} GeV fluxes, Crab, Vela, B1706$-$44 and Geminga, have already been detected by VHE telescopes below 100 GeV.  On the other hand J0218+4232, with $\gamma$-ray flux a factor of 20-50 times lower, has not been detected below 100 GeV by MAGIC.  In this regard, \textit{Fermi} pulsars with high fluxes that are good targets for VHE telescopes are PSR J0007+7303 (CTA1), PSR J2021+3651 and J2021+4026.  Detection of the pair SSC component will be much more sensitive to intrinsic pulsar parameters, since high pair multiplicity and $B_{\rm LC}$ are required.  Consequently, Crab-like pulsars such as B0540$-$69 and B1509$-$58 are the best prospects for detection of the pair SSC.  We argued that the pair SSC component has already been detected from the Crab by MAGIC.  As we discussed in \S \ref{sec:J0218}, MSPs will not be good targets for detection of this component since their pair energies are too high.  Finally, the primary ICS component above 100 GeV is stronger for pulsars with pair spectra extending to low energies, producing high levels of pair SR, and smaller light cylinder radii.  Based on these requirements,  both Crab-like pulsars and Vela-like pulsars with smaller periods will offer the best prospects for detection of this component.  We predict that the primary ICS emission of the Crab should be presently detectable by HAWC and MAGIC.  
We have noted that there is a predicted range of spectral fluxes, depending on selected model parameters. Thus, continued observations and modelling will allow us to constrain parameter degeneracies.

Our modeling of pulsar broadband spectra could certainly be improved in several respects.  It would be more accurate to use the $E_\parallel$ distribution in the magnetosphere of dissipative models, such as those discussed in \citet{Kalapotharakos2018}.  These models show that $E_\parallel$ is not only a function of altitude but also of azimuth around the current sheet and of distance above the current sheet.  The pair spectra could be computed more accurately by performing (at least) 2D simulations of the time-dependent acceleration gaps to determine both the time-averaged energy distribution and multiplicity of particles escaping the gap that seed the full cascade.  This would eliminate the pair multiplicity as a free parameter and produce more accurate pair spectra.  The photon-photon pair attenuation could be treated more accurately by using the stored SR emissivities, used now for the SSC and ICS emission, for the soft-photon densities.  Ultimately, the field and particle distributions from PIC simulations could be used to model the whole broadband spectrum.  At present, this option is not feasible for several reasons.  First, scaling the low part of the particle energy spectrum of the PIC particle energies (that are already artificially low) up to those of real pulsars is problematic.  Second, the current dynamic range of particle energies that can be treated in PIC models is very small (only a few decades), so that most low-energy particles are in a thermal distribution.   Third, the PIC simulations do not have the resolution to treat the PC pair cascades self-consistently so the pair spectrum from local simulations would still need to be injected separately.  Nevertheless, we plan in the future to add the improvements that are most feasible.

\begin{acknowledgments}
We thank Slavko Bogdanov for supplying the \textit{XMM}-Newton and \textit{NuSTAR} data points for PSR J0218+4232 and Alessia Spolon for supplying \textit{Fermi} and MAGIC data points for J0218+4232.  
Resources supporting this work were provided by the NASA High-End Computing (HEC) Program through the NCCS at Goddard Space Flight Center.   We thank Craig Pelissier of the NCCS in particular for help with parallel processing. This work is based on the research supported wholly / in part by the National Research Foundation of South Africa (NRF; Grant Numbers 87613, 90822, 92860, 93278, and 99072). The Grantholder acknowledges that opinions, findings and conclusions or recommendations expressed in any publication generated by the NRF-supported research is that of the author(s), and that the NRF accepts no liability whatsoever in this regard. 
\end{acknowledgments}

\bibliography{VHE}{}

\begin{thebibliography}{}
\expandafter\ifx\csname natexlab\endcsname\relax\def\natexlab#1{#1}\fi
\providecommand{\url}[1]{\href{#1}{#1}}
\providecommand{\dodoi}[1]{doi:~\href{http://doi.org/#1}{\nolinkurl{#1}}}
\providecommand{\doeprint}[1]{\href{http://ascl.net/#1}{\nolinkurl{http://ascl.net/#1}}}
\providecommand{\doarXiv}[1]{\href{https://arxiv.org/abs/#1}{\nolinkurl{https://arxiv.org/abs/#1}}}

\bibitem[{{Abdalla} {et~al.}(2018){Abdalla}, {Aharonian}, {Ait Benkhali},
  {Ang{\"u}ner}, {Arakawa}, {Arcaro}, {Armand}, {Arrieta}, {Backes}, {Barnard},
  {Becherini}, {Becker Tjus}, {Berge}, {Bernhard}, {Bernl{\"o}hr}, {Blackwell},
  {B{\"o}ttcher}, {Boisson}, {Bolmont}, {Bonnefoy}, {Bordas}, {Bregeon},
  {Brun}, {Brun}, {Bryan}, {B{\"u}chele}, {Bulik}, {Bylund}, {Capasso},
  {Caroff}, {Carosi}, {Casanova}, {Cerruti}, {Chakraborty}, {Chandra},
  {Chaves}, {Chen}, {Colafrancesco}, {Condon}, {Davids}, {Deil}, {Devin},
  {deWilt}, {Dirson}, {Djannati-Ata{\"\i}}, {Dmytriiev}, {Donath},
  {Doroshenko}, {Drury}, {Dyks}, {Egberts}, {Emery}, {Ernenwein}, {Eschbach},
  {Fegan}, {Fiasson}, {Fontaine}, {Funk}, {F{\"u}{\ss}ling}, {Gabici},
  {Gallant}, {Gat{\'e}}, {Giavitto}, {Glawion}, {Glicenstein}, {Gottschall},
  {Grondin}, {Hahn}, {Haupt}, {Heinzelmann}, {Henri}, {Hermann}, {Hinton},
  {Hofmann}, {Hoischen}, {Holch}, {Holler}, {Horns}, {Huber}, {Iwasaki},
  {Jacholkowska}, {Jamrozy}, {Jankowsky}, {Jankowsky}, {Jouvin},
  {Jung-Richardt}, {Kastendieck}, {Katarzy{\'n}ski}, {Katsuragawa}, {Katz},
  {Kerszberg}, {Khangulyan}, {Kh{\'e}lifi}, {King}, {Klepser}, {Klu{\'z}niak},
  {Komin}, {Kosack}, {Krakau}, {Kraus}, {Kr{\"u}ger}, {Lamanna}, {Lau},
  {Lefaucheur}, {Lemi{\`e}re}, {Lemoine-Goumard}, {Lenain}, {Leser}, {Lohse},
  {Lorentz}, {L{\'o}pez-Coto}, {Lypova}, {Malyshev}, {Marandon}, {Marcowith},
  {Mariaud}, {Mart{\'\i}-Devesa}, {Marx}, {Maurin}, {Meintjes}, {Mitchell},
  {Moderski}, {Mohamed}, {Mohrmann}, {Moulin}, {Murach}, {Nakashima}, {de
  Naurois}, {Ndiyavala}, {Niederwanger}, {Niemiec}, {Oakes}, {O'Brien},
  {Odaka}, {Ohm}, {Ostrowski}, {Oya}, {Padovani}, {Panter}, {Parsons},
  {Perennes}, {Petrucci}, {Peyaud}, {Piel}, {Pita}, {Poireau}, {Priyana Noel},
  {Prokhorov}, {Prokoph}, {P{\"u}hlhofer}, {Punch}, {Quirrenbach}, {Raab},
  {Rauth}, {Reimer}, {Reimer}, {Renaud}, {Rieger}, {Rinchiuso}, {Romoli},
  {Rowell}, {Rudak}, {Ruiz-Velasco}, {Sahakian}, {Saito}, {Sanchez},
  {Santangelo}, {Sasaki}, {Schlickeiser}, {Sch{\"u}ssler}, {Schulz},
  {Schwanke}, {Schwemmer}, {Seglar-Arroyo}, {Senniappan}, {Seyffert}, {Shafi},
  {Shilon}, {Shiningayamwe}, {Simoni}, {Sinha}, {Sol}, {Spanier}, {Specovius},
  {Spir-Jacob}, {Stawarz}, {Steenkamp}, {Stegmann}, {Steppa}, {Takahashi},
  {Tavernet}, {Tavernier}, {Taylor}, {Terrier}, {Tibaldo}, {Tiziani},
  {Tluczykont}, {Trichard}, {Tsirou}, {Tsuji}, {Tuffs}, {Uchiyama}, {van der
  Walt}, {van Eldik}, {van Rensburg}, {van Soelen}, {Vasileiadis}, {Veh},
  {Venter}, {Vincent}, {Vink}, {Voisin}, {V{\"o}lk}, {Vuillaume}, {Wadiasingh},
  {Wagner}, {Wagner}, {White}, {Wierzcholska}, {Yang}, {Zaborov}, {Zacharias},
  {Zanin}, {Zdziarski}, {Zech}, {Zefi}, {Ziegler}, {Zorn}, {{\.Z}ywucka},
  {Kerr}, {Johnston}, \& {Shannon}}]{Abdalla2018}
{Abdalla}, H., {Aharonian}, F., {Ait Benkhali}, F., {et~al.} 2018, \aap, 620,
  A66, \dodoi{10.1051/0004-6361/201732153}

\bibitem[{{Abdo} {et~al.}(2010{\natexlab{a}}){Abdo}, {Ackermann}, {Ajello},
  {Atwood}, {Axelsson}, {Baldini}, {Ballet}, {Barbiellini}, {Baring},
  {Bastieri}, {Baughman}, {Bechtol}, {Bellazzini}, {Berenji}, {Blandford},
  {Bloom}, {Bonamente}, {Borgland}, {Bregeon}, {Brez}, {Brigida}, {Bruel},
  {Burnett}, {Buson}, {Caliandro}, {Cameron}, {Camilo}, {Caraveo},
  {Casandjian}, {Cecchi}, {{\c{C}}elik}, {Charles}, {Chekhtman}, {Cheung},
  {Chiang}, {Ciprini}, {Claus}, {Cognard}, {Cohen-Tanugi}, {Cominsky},
  {Conrad}, {Corbet}, {Cutini}, {den Hartog}, {Dermer}, {de Angelis}, {de
  Luca}, {de Palma}, {Digel}, {Dormody}, {Silva}, {Drell}, {Dubois}, {Dumora},
  {Espinoza}, {Farnier}, {Favuzzi}, {Fegan}, {Ferrara}, {Focke}, {Fortin},
  {Frailis}, {Freire}, {Fukazawa}, {Funk}, {Fusco}, {Gargano}, {Gasparrini},
  {Gehrels}, {Germani}, {Giavitto}, {Giebels}, {Giglietto}, {Giommi},
  {Giordano}, {Glanzman}, {Godfrey}, {Gotthelf}, {Grenier}, {Grondin}, {Grove},
  {Guillemot}, {Guiriec}, {Gwon}, {Hanabata}, {Harding}, {Hayashida}, {Hays},
  {Hughes}, {Jackson}, {J{\'o}hannesson}, {Johnson}, {Johnson}, {Johnson},
  {Johnson}, {Johnston}, {Kamae}, {Kanbach}, {Kaspi}, {Katagiri}, {Kataoka},
  {Kawai}, {Kerr}, {Kn{\"o}dlseder}, {Kocian}, {Kramer}, {Kuss}, {Lande},
  {Latronico}, {Lemoine-Goumard}, {Livingstone}, {Longo}, {Loparco}, {Lott},
  {Lovellette}, {Lubrano}, {Lyne}, {Madejski}, {Makeev}, {Manchester},
  {Marelli}, {Mazziotta}, {McConville}, {McEnery}, {McGlynn}, {Meurer},
  {Michelson}, {Mineo}, {Mitthumsiri}, {Mizuno}, {Moiseev}, {Monte}, {Monzani},
  {Morselli}, {Moskalenko}, {Murgia}, {Nakamori}, {Nolan}, {Norris}, {Noutsos},
  {Nuss}, {Ohsugi}, {Omodei}, {Orlando}, {Ormes}, {Ozaki}, {Paneque},
  {Panetta}, {Parent}, {Pelassa}, {Pepe}, {Pesce-Rollins}, {Piron}, {Porter},
  {Rain{\`o}}, {Rando}, {Ransom}, {Ray}, {Razzano}, {Rea}, {Reimer}, {Reimer},
  {Reposeur}, {Ritz}, {Rodriguez}, {Romani}, {Roth}, {Ryde}, {Sadrozinski},
  {Sanchez}, {Sander}, {Saz Parkinson}, {Scargle}, {Schalk}, {Sellerholm},
  {Sgr{\`o}}, {Siskind}, {Smith}, {Smith}, {Spandre}, {Spinelli}, {Stappers},
  {Starck}, {Striani}, {Strickman}, {Strong}, {Suson}, {Tajima}, {Takahashi},
  {Takahashi}, {Tanaka}, {Thayer}, {Thayer}, {Theureau}, {Thompson},
  {Thorsett}, {Tibaldo}, {Tibolla}, {Torres}, {Tosti}, {Tramacere}, {Uchiyama},
  {Usher}, {Van Etten}, {Vasileiou}, {Venter}, {Vilchez}, {Vitale}, {Waite},
  {Wang}, {Wang}, {Watters}, {Weltevrede}, {Winer}, {Wood}, {Ylinen}, \&
  {Ziegler}}]{Abdo2010PC1}
{Abdo}, A.~A., {Ackermann}, M., {Ajello}, M., {et~al.} 2010{\natexlab{a}},
  \apjs, 187, 460, \dodoi{10.1088/0067-0049/187/2/460}

\bibitem[{{Abdo} {et~al.}(2010{\natexlab{b}}){Abdo}, {Ackermann}, {Ajello},
  {Allafort}, {Atwood}, {Baldini}, {Ballet}, {Barbiellini}, {Baring},
  {Bastieri}, {Baughman}, {Bechtol}, {Bellazzini}, {Berenji}, {Blandford},
  {Bloom}, {Bonamente}, {Borgland}, {Bouvier}, {Bregeon}, {Brez}, {Brigida},
  {Bruel}, {Burnett}, {Buson}, {Caliandro}, {Cameron}, {Caraveo}, {Carrigan},
  {Casand jian}, {Cecchi}, {{\c{C}}elik}, {Chekhtman}, {Cheung}, {Chiang},
  {Ciprini}, {Claus}, {Cohen-Tanugi}, {Conrad}, {Dermer}, {de Luca}, {de
  Palma}, {Dormody}, {Silva}, {Drell}, {Dubois}, {Dumora}, {Farnier},
  {Favuzzi}, {Fegan}, {Focke}, {Fortin}, {Frailis}, {Fukazawa}, {Funk},
  {Fusco}, {Gargano}, {Gasparrini}, {Gehrels}, {Germani}, {Giavitto},
  {Giebels}, {Giglietto}, {Giordano}, {Glanzman}, {Godfrey}, {Grenier},
  {Grondin}, {Grove}, {Guillemot}, {Guiriec}, {Hadasch}, {Harding}, {Hays},
  {Hobbs}, {Horan}, {Hughes}, {Jackson}, {J{\'o}hannesson}, {Johnson},
  {Johnson}, {Johnson}, {Kamae}, {Katagiri}, {Kataoka}, {Kawai}, {Kerr},
  {Kn{\"o}dlseder}, {Kuss}, {Lande}, {Latronico}, {Lee}, {Lemoine-Goumard},
  {Llena Garde}, {Longo}, {Loparco}, {Lott}, {Lovellette}, {Lubrano}, {Makeev},
  {Manchester}, {Marelli}, {Mazziotta}, {McConville}, {McEnery}, {McGlynn},
  {Meurer}, {Michelson}, {Mitthumsiri}, {Mizuno}, {Moiseev}, {Monte},
  {Monzani}, {Morselli}, {Moskalenko}, {Murgia}, {Nakamori}, {Nolan}, {Norris},
  {Noutsos}, {Nuss}, {Ohsugi}, {Omodei}, {Orland o}, {Ormes}, {Ozaki},
  {Paneque}, {Panetta}, {Parent}, {Pelassa}, {Pepe}, {Pesce-Rollins},
  {Pierbattista}, {Piron}, {Porter}, {Rain{\`o}}, {Rando}, {Ray}, {Razzano},
  {Reimer}, {Reimer}, {Reposeur}, {Ritz}, {Rochester}, {Rodriguez}, {Romani},
  {Roth}, {Ryde}, {Sadrozinski}, {Sander}, {Saz Parkinson}, {Sgr{\`o}},
  {Siskind}, {Smith}, {Smith}, {Spandre}, {Spinelli}, {Starck}, {Strickman},
  {Suson}, {Takahashi}, {Takahashi}, {Tanaka}, {Thayer}, {Thayer}, {Thompson},
  {Tibaldo}, {Torres}, {Tosti}, {Tramacere}, {Usher}, {Van Etten}, {Vasileiou},
  {Venter}, {Vilchez}, {Vitale}, {Waite}, {Wang}, {Watters}, {Weltevrede},
  {Winer}, {Wood}, {Ylinen}, \& {Ziegler}}]{Abdo2010Vela}
---. 2010{\natexlab{b}}, \apj, 713, 154, \dodoi{10.1088/0004-637X/713/1/154}

\bibitem[{{Abdo} {et~al.}(2010{\natexlab{c}}){Abdo}, {Ackermann}, {Ajello},
  {Baldini}, {Ballet}, {Barbiellini}, {Bastieri}, {Baughman}, {Bechtol},
  {Bellazzini}, {Berenji}, {Bignami}, {Blandford}, {Bloom}, {Bonamente},
  {Borgland}, {Bregeon}, {Brez}, {Brigida}, {Bruel}, {Burnett}, {Caliandro},
  {Cameron}, {Caraveo}, {Casandjian}, {Cecchi}, {{\c{C}}elik}, {Charles},
  {Chekhtman}, {Cheung}, {Chiang}, {Ciprini}, {Claus}, {Cohen-Tanugi},
  {Conrad}, {Dermer}, {de Palma}, {Dormody}, {Silva}, {Drell}, {Dubois},
  {Dumora}, {Edmonds}, {Farnier}, {Favuzzi}, {Fegan}, {Focke}, {Fortin},
  {Frailis}, {Fukazawa}, {Funk}, {Fusco}, {Gargano}, {Gasparrini}, {Gehrels},
  {Germani}, {Giavitto}, {Giglietto}, {Giordano}, {Glanzman}, {Godfrey},
  {Grenier}, {Grondin}, {Grove}, {Guillemot}, {Guiriec}, {Hadasch}, {Harding},
  {Hays}, {Hughes}, {J{\'o}hannesson}, {Johnson}, {Johnson}, {Johnson},
  {Kamae}, {Katagiri}, {Kataoka}, {Kawai}, {Kerr}, {Kn{\"o}dlseder}, {Kuss},
  {Lande}, {Latronico}, {Lemoine-Goumard}, {Longo}, {Loparco}, {Lott},
  {Lovellette}, {Lubrano}, {Makeev}, {Marelli}, {Mazziotta}, {McEnery},
  {Meurer}, {Michelson}, {Mitthumsiri}, {Mizuno}, {Moiseev}, {Monte},
  {Monzani}, {Morselli}, {Moskalenko}, {Murgia}, {Nolan}, {Norris}, {Nuss},
  {Ohsugi}, {Omodei}, {Orlando}, {Ormes}, {Ozaki}, {Paneque}, {Panetta},
  {Parent}, {Pelassa}, {Pepe}, {Pesce-Rollins}, {Piron}, {Porter}, {Rain{\`o}},
  {Rando}, {Ray}, {Razzano}, {Reimer}, {Reimer}, {Reposeur}, {Rochester},
  {Rodriguez}, {Romani}, {Roth}, {Ryde}, {Sadrozinski}, {Sand er}, {Saz
  Parkinson}, {Scargle}, {Sgr{\`o}}, {Siskind}, {Smith}, {Smith}, {Spandre},
  {Spinelli}, {Strickman}, {Suson}, {Takahashi}, {Takahashi}, {Tanaka},
  {Thayer}, {Thayer}, {Thompson}, {Tibaldo}, {Torres}, {Tosti}, {Tramacere},
  {Usher}, {Van Etten}, {Vasileiou}, {Venter}, {Vilchez}, {Vitale}, {Waite},
  {Wang}, {Watters}, {Winer}, {Wood}, {Ylinen}, \& {Ziegler}}]{Abdo2010Geminga}
---. 2010{\natexlab{c}}, \apj, 720, 272, \dodoi{10.1088/0004-637X/720/1/272}

\bibitem[{{Abdo} {et~al.}(2013){Abdo}, {Ajello}, {Allafort}, {Baldini},
  {Ballet}, {Barbiellini}, {Baring}, {Bastieri}, {Belfiore}, {Bellazzini},
  {Bhattacharyya}, {Bissaldi}, {Bloom}, {Bonamente}, {Bottacini}, {Brandt},
  {Bregeon}, {Brigida}, {Bruel}, {Buehler}, {Burgay}, {Burnett}, {Busetto},
  {Buson}, {Caliandro}, {Cameron}, {Camilo}, {Caraveo}, {Casandjian}, {Cecchi},
  {{\c{C}}elik}, {Charles}, {Chaty}, {Chaves}, {Chekhtman}, {Chen}, {Chiang},
  {Chiaro}, {Ciprini}, {Claus}, {Cognard}, {Cohen-Tanugi}, {Cominsky},
  {Conrad}, {Cutini}, {D'Ammando}, {de Angelis}, {DeCesar}, {De Luca}, {den
  Hartog}, {de Palma}, {Dermer}, {Desvignes}, {Digel}, {Di Venere}, {Drell},
  {Drlica-Wagner}, {Dubois}, {Dumora}, {Espinoza}, {Falletti}, {Favuzzi},
  {Ferrara}, {Focke}, {Franckowiak}, {Freire}, {Funk}, {Fusco}, {Gargano},
  {Gasparrini}, {Germani}, {Giglietto}, {Giommi}, {Giordano}, {Giroletti},
  {Glanzman}, {Godfrey}, {Gotthelf}, {Grenier}, {Grondin}, {Grove},
  {Guillemot}, {Guiriec}, {Hadasch}, {Hanabata}, {Harding}, {Hayashida},
  {Hays}, {Hessels}, {Hewitt}, {Hill}, {Horan}, {Hou}, {Hughes}, {Jackson},
  {Janssen}, {Jogler}, {J{\'o}hannesson}, {Johnson}, {Johnson}, {Johnson},
  {Johnson}, {Johnston}, {Kamae}, {Kataoka}, {Keith}, {Kerr}, {Kn{\"o}dlseder},
  {Kramer}, {Kuss}, {Lande}, {Larsson}, {Latronico}, {Lemoine-Goumard},
  {Longo}, {Loparco}, {Lovellette}, {Lubrano}, {Lyne}, {Manchester}, {Marelli},
  {Massaro}, {Mayer}, {Mazziotta}, {McEnery}, {McLaughlin}, {Mehault},
  {Michelson}, {Mignani}, {Mitthumsiri}, {Mizuno}, {Moiseev}, {Monzani},
  {Morselli}, {Moskalenko}, {Murgia}, {Nakamori}, {Nemmen}, {Nuss}, {Ohno},
  {Ohsugi}, {Orienti}, {Orlando}, {Ormes}, {Paneque}, {Panetta}, {Parent},
  {Perkins}, {Pesce-Rollins}, {Pierbattista}, {Piron}, {Pivato}, {Pletsch},
  {Porter}, {Possenti}, {Rain{\`o}}, {Rando}, {Ransom}, {Ray}, {Razzano},
  {Rea}, {Reimer}, {Reimer}, {Renault}, {Reposeur}, {Ritz}, {Romani}, {Roth},
  {Rousseau}, {Roy}, {Ruan}, {Sartori}, {Saz Parkinson}, {Scargle}, {Schulz},
  {Sgr{\`o}}, {Shannon}, {Siskind}, {Smith}, {Spandre}, {Spinelli}, {Stappers},
  {Strong}, {Suson}, {Takahashi}, {Thayer}, {Thayer}, {Theureau}, {Thompson},
  {Thorsett}, {Tibaldo}, {Tibolla}, {Tinivella}, {Torres}, {Tosti}, {Troja},
  {Uchiyama}, {Usher}, {Vandenbroucke}, {Vasileiou}, {Venter}, {Vianello},
  {Vitale}, {Wang}, {Weltevrede}, {Winer}, {Wolff}, {Wood}, {Wood}, {Wood}, \&
  {Yang}}]{Abdo2013}
{Abdo}, A.~A., {Ajello}, M., {Allafort}, A., {et~al.} 2013, \apjs, 208, 17,
  \dodoi{10.1088/0067-0049/208/2/17}

\bibitem[{{Abeysekara} {et~al.}(2017){Abeysekara}, {Albert}, {Alfaro},
  {Alvarez}, {{\'A}lvarez}, {Arceo}, {Arteaga-Vel{\'a}zquez}, {Ayala Solares},
  {Barber}, {Bautista-Elivar}, {Becerril}, {Belmont-Moreno}, {BenZvi},
  {Berley}, {Braun}, {Brisbois}, {Caballero-Mora}, {Capistr{\'a}n},
  {Carrami{\~n}ana}, {Casanova}, {Castillo}, {Cotti}, {Cotzomi}, {Couti{\~n}o
  de Le{\'o}n}, {de la Fuente}, {De Le{\'o}n}, {DeYoung}, {Dingus},
  {DuVernois}, {D{\'\i}az-V{\'e}lez}, {Ellsworth}, {Fiorino}, {Fraija},
  {Garc{\'\i}a-Gonz{\'a}lez}, {Gerhardt}, {Gonz{\'a}lez Mun{\"o}z},
  {Gonz{\'a}lez}, {Goodman}, {Hampel-Arias}, {Harding}, {Hernandez},
  {Hernandez-Almada}, {Hinton}, {Hui}, {H{\"u}ntemeyer}, {Iriarte},
  {Jardin-Blicq}, {Joshi}, {Kaufmann}, {Kieda}, {Lara}, {Lauer}, {Lee},
  {Lennarz}, {Le{\'o}n Vargas}, {Linnemann}, {Longinotti}, {Raya},
  {Luna-Garc{\'\i}a}, {L{\'o}pez-Coto}, {Malone}, {Marinelli}, {Martinez},
  {Martinez-Castellanos}, {Mart{\'\i}nez-Castro}, {Mart{\'\i}nez-Huerta},
  {Matthews}, {Miranda-Romagnoli}, {Moreno}, {Mostaf{\'a}}, {Nellen},
  {Newbold}, {Nisa}, {Noriega-Papaqui}, {Pelayo}, {Pretz},
  {P{\'e}rez-P{\'e}rez}, {Ren}, {Rho}, {Rivi{\`e}re}, {Rosa-Gonz{\'a}lez},
  {Rosenberg}, {Ruiz-Velasco}, {Salazar}, {Salesa Greus}, {Sandoval},
  {Schneider}, {Schoorlemmer}, {Sinnis}, {Smith}, {Springer}, {Surajbali},
  {Taboada}, {Tibolla}, {Tollefson}, {Torres}, {Ukwatta}, {Villase{\~n}or},
  {Weisgarber}, {Westerhoff}, {Wisher}, {Wood}, {Yapici}, {Yodh}, {Younk},
  {Zepeda}, \& {Zhou}}]{Abeysekara2017}
{Abeysekara}, A.~U., {Albert}, A., {Alfaro}, R., {et~al.} 2017, \apj, 843, 39,
  \dodoi{10.3847/1538-4357/aa7555}

\bibitem[{{Acciari} {et~al.}(2020){Acciari}, {Ansoldi}, {Antonelli},
  {Antoranz}, {Babic}, {Banerjee}, {Bangale}, {Barres de Almeida}, {Barrio},
  {Becerra Gonz{\'a}lez}, {Bednarek}, {Bernardini}, {Berti}, {Biasuzzi},
  {Biland}, {Blanch}, {Bonnefoy}, {Bonnoli}, {Borracci}, {Bretz}, {Buson},
  {Carosi}, {Chatterjee}, {Clavero}, {Colin}, {Colombo}, {Contreras},
  {Cortina}, {Covino}, {Da Vela}, {Dazzi}, {De Angelis}, {De Lotto}, {de
  O{\~n}a Wilhelmi}, {Di Pierro}, {Doert}, {Dom{\'\i}nguez}, {Dominis Prester},
  {Dorner}, {Doro}, {Einecke}, {Eisenacher Glawion}, {Elsaesser}, {Fallah
  Ramazani}, {Fern{\'a}ndez-Barral}, {Fidalgo}, {Fonseca}, {Font}, {Frantzen},
  {Fruck}, {Galindo}, {Garc{\'\i}a L{\'o}pez}, {Garczarczyk}, {Garrido
  Terrats}, {Gaug}, {Giammaria}, {Godinovi{\'c}}, {Gonz{\'a}lez Mu{\~n}oz},
  {Gora}, {Guberman}, {Hadasch}, {Hahn}, {Hanabata}, {Hayashida}, {Herrera},
  {Hose}, {Hrupec}, {Hughes}, {Idec}, {Kodani}, {Konno}, {Kubo}, {Kushida}, {La
  Barbera}, {Lelas}, {Lindfors}, {Lombardi}, {Longo}, {L{\'o}pez},
  {L{\'o}pez-Coto}, {Majumdar}, {Makariev}, {Mallot}, {Maneva}, {Manganaro},
  {Mannheim}, {Maraschi}, {Marcote}, {Mariotti}, {Mart{\'\i}nez}, {Mazin},
  {Menzel}, {Mirand a}, {Mirzoyan}, {Moralejo}, {Moretti}, {Nakajima},
  {Neustroev}, {Niedzwiecki}, {Nievas Rosillo}, {Nilsson}, {Nishijima}, {Noda},
  {Nogu{\'e}s}, {Overkemping}, {Paiano}, {Palacio}, {Palatiello}, {Paneque},
  {Paoletti}, {Paredes}, {Paredes-Fortuny}, {Pedaletti}, {Peresano}, {Perri},
  {Persic}, {Poutanen}, {Prada Moroni}, {Prandini}, {Puljak}, {Reichardt},
  {Rhode}, {Rib{\'o}}, {Rico}, {Rodriguez Garcia}, {Saito}, {Satalecka},
  {Schultz}, {Schweizer}, {Shore}, {Sillanp{\"a}{\"a}}, {Sitarek}, {Snidaric},
  {Sobczynska}, {Stamerra}, {Steinbring}, {Strzys}, {Suri{\'c}}, {Takalo},
  {Tavecchio}, {Temnikov}, {Terzi{\'c}}, {Tescaro}, {Teshima}, {Thaele},
  {Torres}, {Toyama}, {Treves}, {Vanzo}, {Verguilov}, {Vovk}, {Ward}, {Will},
  {Wu}, \& {Zanin}}]{Acciari2020}
{Acciari}, V.~A., {Ansoldi}, S., {Antonelli}, L.~A., {et~al.} 2020, \aap, 643,
  L14, \dodoi{10.1051/0004-6361/202039131}

\bibitem[{{Aleksi{\'c}} {et~al.}(2011){Aleksi{\'c}}, {Alvarez}, {Antonelli},
  {Antoranz}, {Asensio}, {Backes}, {Barrio}, {Bastieri}, {Becerra
  Gonz{\'a}lez}, {Bednarek}, {Berdyugin}, {Berger}, {Bernardini}, {Biland},
  {Blanch}, {Bock}, {Boller}, {Bonnoli}, {Borla Tridon}, {Braun}, {Bretz},
  {Ca{\~n}ellas}, {Carmona}, {Carosi}, {Colin}, {Colombo}, {Contreras},
  {Cortina}, {Cossio}, {Covino}, {Dazzi}, {De Angelis}, {De Caneva}, {De Cea
  del Pozo}, {De Lotto}, {Delgado Mendez}, {Diago Ortega}, {Doert},
  {Dom{\'\i}nguez}, {Dominis Prester}, {Dorner}, {Doro}, {Eisenacher},
  {Elsaesser}, {Ferenc}, {Fonseca}, {Font}, {Fruck}, {Garc{\'\i}a L{\'o}pez},
  {Garczarczyk}, {Garrido}, {Giavitto}, {Godinovi{\'c}}, {Hadasch},
  {H{\"a}fner}, {Herrero}, {Hildebrand}, {H{\"o}hne-M{\"o}nch}, {Hose},
  {Hrupec}, {Jogler}, {Kellermann}, {Klepser}, {Kr{\"a}henb{\"u}hl}, {Krause},
  {Kushida}, {La Barbera}, {Lelas}, {Leonardo}, {Lindfors}, {Lombardi},
  {L{\'o}pez}, {L{\'o}pez-Oramas}, {Lorenz}, {Makariev}, {Maneva},
  {Mankuzhiyil}, {Mannheim}, {Maraschi}, {Marcote}, {Mariotti},
  {Mart{\'\i}nez}, {Mazin}, {Meucci}, {Mirand a}, {Mirzoyan}, {Mold{\'o}n},
  {Moralejo}, {Munar-Adrover}, {Nieto}, {Nilsson}, {Orito}, {Otte}, {Oya},
  {Paneque}, {Paoletti}, {Pardo}, {Paredes}, {Partini}, {Perez-Torres},
  {Persic}, {Peruzzo}, {Pilia}, {Pochon}, {Prada}, {Prada Moroni}, {Prandini},
  {Puerto Gimenez}, {Puljak}, {Reichardt}, {Reinthal}, {Rhode}, {Rib{\'o}},
  {Rico}, {Rissi}, {R{\"u}gamer}, {Saggion}, {Saito}, {Saito}, {Salvati},
  {Satalecka}, {Scalzotto}, {Scapin}, {Schultz}, {Schweizer}, {Shayduk},
  {Shore}, {Sillanp{\"a}{\"a}}, {Sitarek}, {Snidaric}, {Sobczynska}, {Spanier},
  {Spiro}, {Stamatescu}, {Stamerra}, {Steinke}, {Storz}, {Strah}, {Suri{\'c}},
  {Takalo}, {Takami}, {Tavecchio}, {Temnikov}, {Terzi{\'c}}, {Tescaro},
  {Teshima}, {Tibolla}, {Torres}, {Treves}, {Uellenbeck}, {Vankov}, {Vogler},
  {Wagner}, {Weitzel}, {Zabalza}, {Zandanel}, {Zanin}, \&
  {Hirotani}}]{Aleksic2011}
{Aleksi{\'c}}, J., {Alvarez}, E.~A., {Antonelli}, L.~A., {et~al.} 2011, \apj,
  742, 43, \dodoi{10.1088/0004-637X/742/1/43}

\bibitem[{{Aleksi{\'c}} {et~al.}(2012){Aleksi{\'c}}, {Alvarez}, {Antonelli},
  {Antoranz}, {Asensio}, {Backes}, {Barrio}, {Bastieri}, {Becerra
  Gonz{\'a}lez}, {Bednarek}, {Berdyugin}, {Berger}, {Bernardini}, {Biland},
  {Blanch}, {Bock}, {Boller}, {Bonnoli}, {Borla Tridon}, {Braun}, {Bretz},
  {Ca{\~n}ellas}, {Carmona}, {Carosi}, {Colin}, {Colombo}, {Contreras},
  {Cortina}, {Cossio}, {Covino}, {Dazzi}, {de Angelis}, {de Caneva}, {de Cea
  Del Pozo}, {de Lotto}, {Delgado Mendez}, {Diago Ortega}, {Doert},
  {Dom{\'\i}nguez}, {Dominis Prester}, {Dorner}, {Doro}, {Eisenacher},
  {Elsaesser}, {Ferenc}, {Fonseca}, {Font}, {Fruck}, {Garc{\'\i}a L{\'o}pez},
  {Garczarczyk}, {Garrido}, {Giavitto}, {Godinovi{\'c}}, {Hadasch},
  {H{\"a}fner}, {Herrero}, {Hildebrand}, {H{\"o}hne-M{\"o}nch}, {Hose},
  {Hrupec}, {Jogler}, {Kellermann}, {Klepser}, {Kr{\"a}henb{\"u}hl}, {Krause},
  {Kushida}, {La Barbera}, {Lelas}, {Leonardo}, {Lewand owska}, {Lindfors},
  {Lombardi}, {L{\'o}pez}, {L{\'o}pez-Oramas}, {Lorenz}, {Makariev}, {Maneva},
  {Mankuzhiyil}, {Mannheim}, {Maraschi}, {Mariotti}, {Mart{\'\i}nez}, {Mazin},
  {Meucci}, {Miranda}, {Mirzoyan}, {Mold{\'o}n}, {Moralejo}, {Munar-Adrover},
  {Niedzwiecki}, {Nieto}, {Nilsson}, {Nowak}, {Orito}, {Paneque}, {Paoletti},
  {Pardo}, {Paredes}, {Partini}, {Perez-Torres}, {Persic}, {Peruzzo}, {Pilia},
  {Pochon}, {Prada}, {Prada Moroni}, {Prandini}, {Puerto Gimenez}, {Puljak},
  {Reichardt}, {Reinthal}, {Rhode}, {Rib{\'o}}, {Rico}, {R{\"u}gamer},
  {Saggion}, {Saito}, {Saito}, {Salvati}, {Satalecka}, {Scalzotto}, {Scapin},
  {Schultz}, {Schweizer}, {Shayduk}, {Shore}, {Sillanp{\"a}{\"a}}, {Sitarek},
  {{\v{S}}nidari{\'c}}, {Sobczynska}, {Spanier}, {Spiro}, {Stamatescu},
  {Stamerra}, {Steinke}, {Storz}, {Strah}, {Suri{\'c}}, {Takalo}, {Takami},
  {Tavecchio}, {Temnikov}, {Terzi{\'c}}, {Tescaro}, {Teshima}, {Tibolla},
  {Torres}, {Treves}, {Uellenbeck}, {Vankov}, {Vogler}, {Wagner}, {Weitzel},
  {Zabalza}, {Zandanel}, {Zanin}, \& {Hirotani}}]{Aleksic2012}
---. 2012, \aap, 540, A69, \dodoi{10.1051/0004-6361/201118166}

\bibitem[{{Aliu} {et~al.}(2008){Aliu}, {Anderhub}, {Antonelli}, {Antoranz},
  {Backes}, {Baixeras}, {Barrio}, {Bartko}, {Bastieri}, {Becker}, {Bednarek},
  {Berger}, {Bernardini}, {Bigongiari}, {Biland}, {Bock}, {Bonnoli}, {Bordas},
  {Bosch-Ramon}, {Bretz}, {Britvitch}, {Camara}, {Carmona}, {Chilingarian},
  {Commichau}, {Contreras}, {Cortina}, {Costado}, {Covino}, {Curtef}, {Dazzi},
  {De Angelis}, {De Cea del Pozo}, {de los Reyes}, {De Lotto}, {De Maria}, {De
  Sabata}, {Delgado Mendez}, {Dominguez}, {Dorner}, {Doro}, {Els{\"a}sser},
  {Errando}, {Fagiolini}, {Ferenc}, {Fernandez}, {Firpo}, {Fonseca}, {Font},
  {Galante}, {Garcia Lopez}, {Garczarczyk}, {Gaug}, {Goebel}, {Hadasch},
  {Hayashida}, {Herrero}, {H{\"o}hne}, {Hose}, {Hsu}, {Huber}, {Jogler},
  {Kranich}, {La Barbera}, {Laille}, {Leonardo}, {Lindfors}, {Lombardi},
  {Longo}, {Lopez}, {Lorenz}, {Majumdar}, {Maneva}, {Mankuzhiyil}, {Mannheim},
  {Maraschi}, {Mariotti}, {Martinez}, {Mazin}, {Meucci}, {Meyer}, {Miranda},
  {Mirzoyan}, {Moles}, {Moralejo}, {Nieto}, {Nilsson}, {Ninkovic}, {Otte},
  {Oya}, {Paoletti}, {Paredes}, {Pasanen}, {Pascoli}, {Pauss}, {Pegna},
  {Perez-Torres}, {Persic}, {Peruzzo}, {Piccioli}, {Prada}, {Prandini},
  {Puchades}, {Raymers}, {Rhode}, {Rib{\'o}}, {Rico}, {Rissi}, {Robert},
  {R{\"u}gamer}, {Saggion}, {Saito}, {Salvati}, {Sanchez-Conde}, {Sartori},
  {Satalecka}, {Scalzotto}, {Scapin}, {Schweizer}, {Shayduk}, {Shinozaki},
  {Shore}, {Sidro}, {Sierpowska-Bartosik}, {Sillanp{\"a}{\"a}}, {Sobczynska},
  {Spanier}, {Stamerra}, {Stark}, {Takalo}, {Tavecchio}, {Temnikov}, {Tescaro},
  {Teshima}, {Tluczykont}, {Torres}, {Turini}, {Vankov}, {Venturini}, {Vitale},
  {Wagner}, {Wittek}, {Zabalza}, {Zandanel}, {Zanin}, {Zapatero}, {de Jager},
  {de Ona Wilhelmi}, \& {MAGIC Collaboration}}]{Aliu2008}
{Aliu}, E., {Anderhub}, H., {Antonelli}, L.~A., {et~al.} 2008, Science, 322,
  1221, \dodoi{10.1126/science.1164718}

\bibitem[{{Aliu} {et~al.}(2011){Aliu}, {Arlen}, {Aune}, {Beilicke}, {Benbow},
  {Bouvier}, {Bradbury}, {Buckley}, {Bugaev}, {Byrum}, {Cannon}, {Cesarini},
  {Christiansen}, {Ciupik}, {Collins-Hughes}, {Connolly}, {Cui}, {Dickherber},
  {Duke}, {Errando}, {Falcone}, {Finley}, {Finnegan}, {Fortson}, {Furniss},
  {Galante}, {Gall}, {Gibbs}, {Gillanders}, {Godambe}, {Griffin}, {Grube},
  {Guenette}, {Gyuk}, {Hanna}, {Holder}, {Huan}, {Hughes}, {Hui}, {Humensky},
  {Imran}, {Kaaret}, {Karlsson}, {Kertzman}, {Kieda}, {Krawczynski},
  {Krennrich}, {Lang}, {Lyutikov}, {Madhavan}, {Maier}, {Majumdar}, {McArthur},
  {McCann}, {McCutcheon}, {Moriarty}, {Mukherjee}, {Nu{\~n}ez}, {Ong}, {Orr},
  {Otte}, {Park}, {Perkins}, {Pizlo}, {Pohl}, {Prokoph}, {Quinn}, {Ragan},
  {Reyes}, {Reynolds}, {Roache}, {Rose}, {Ruppel}, {Saxon}, {Schroedter},
  {Sembroski}, {{\c{S}}ent{\"u}rk}, {Smith}, {Staszak}, {Te{\v{s}}i{\'c}},
  {Theiling}, {Thibadeau}, {Tsurusaki}, {Tyler}, {Varlotta}, {Vassiliev},
  {Vincent}, {Vivier}, {Wakely}, {Ward}, {Weekes}, {Weinstein}, {Weisgarber},
  {Williams}, \& {Zitzer}}]{Aliu2011}
{Aliu}, E., {Arlen}, T., {Aune}, T., {et~al.} 2011, Science, 334, 69,
  \dodoi{10.1126/science.1208192}

\bibitem[{{Ansoldi} {et~al.}(2016){Ansoldi}, {Antonelli}, {Antoranz}, {Babic},
  {Bangale}, {Barres de Almeida}, {Barrio}, {Becerra Gonz{\'a}lez}, {Bednarek},
  {Bernardini}, {Biasuzzi}, {Biland}, {Blanch}, {Bonnefoy}, {Bonnoli},
  {Borracci}, {Bretz}, {Carmona}, {Carosi}, {Colin}, {Colombo}, {Contreras},
  {Cortina}, {Covino}, {Da Vela}, {Dazzi}, {De Angelis}, {De Caneva}, {De
  Lotto}, {de O{\~n}a Wilhelmi}, {Delgado Mendez}, {Di Pierro}, {Dominis
  Prester}, {Dorner}, {Doro}, {Einecke}, {Eisenacher Glawion}, {Elsaesser},
  {Fern{\'a}ndez-Barral}, {Fidalgo}, {Fonseca}, {Font}, {Frantzen}, {Fruck},
  {Galindo}, {Garc{\'\i}a L{\'o}pez}, {Garczarczyk}, {Garrido Terrats}, {Gaug},
  {Godinovi{\'c}}, {Gonz{\'a}lez Mu{\~n}oz}, {Gozzini}, {Hanabata},
  {Hayashida}, {Herrera}, {Hirotani}, {Hose}, {Hrupec}, {Hughes}, {Idec},
  {Kellermann}, {Knoetig}, {Kodani}, {Konno}, {Krause}, {Kubo}, {Kushida}, {La
  Barbera}, {Lelas}, {Lewandowska}, {Lindfors}, {Lombardi}, {Longo},
  {L{\'o}pez}, {L{\'o}pez-Coto}, {L{\'o}pez-Oramas}, {Lorenz}, {Makariev},
  {Mallot}, {Maneva}, {Mannheim}, {Maraschi}, {Marcote}, {Mariotti},
  {Mart{\'\i}nez}, {Mazin}, {Menzel}, {Miranda}, {Mirzoyan}, {Moralejo},
  {Munar-Adrover}, {Nakajima}, {Neustroev}, {Niedzwiecki}, {Nevas Rosillo},
  {Nilsson}, {Nishijima}, {Noda}, {Orito}, {Overkemping}, {Paiano},
  {Palatiello}, {Paneque}, {Paoletti}, {Paredes}, {Paredes-Fortuny}, {Persic},
  {Poutanen}, {Prada Moroni}, {Prandini}, {Puljak}, {Reinthal}, {Rhode},
  {Rib{\'o}}, {Rico}, {Rodriguez Garcia}, {Saito}, {Saito}, {Satalecka},
  {Scalzotto}, {Scapin}, {Schultz}, {Schweizer}, {Shore}, {Sillanp{\"a}{\"a}},
  {Sitarek}, {Snidaric}, {Sobczynska}, {Stamerra}, {Steinbring}, {Strzys},
  {Takalo}, {Takami}, {Tavecchio}, {Temnikov}, {Terzi{\'c}}, {Tescaro},
  {Teshima}, {Thaele}, {Torres}, {Toyama}, {Treves}, {Ward}, {Will}, \&
  {Zanin}}]{Ansoldi2016}
{Ansoldi}, S., {Antonelli}, L.~A., {Antoranz}, P., {et~al.} 2016, \aap, 585,
  A133, \dodoi{10.1051/0004-6361/201526853}

\bibitem[{{Atwood} {et~al.}(2009){Atwood}, {Abdo}, {Ackermann}, {Althouse},
  {Anderson}, {Axelsson}, {Baldini}, {Ballet}, {Band}, {Barbiellini},
  {Bartelt}, {Bastieri}, {Baughman}, {Bechtol}, {B{\'e}d{\'e}r{\`e}de},
  {Bellardi}, {Bellazzini}, {Berenji}, {Bignami}, {Bisello}, {Bissaldi},
  {Blandford}, {Bloom}, {Bogart}, {Bonamente}, {Bonnell}, {Borgland },
  {Bouvier}, {Bregeon}, {Brez}, {Brigida}, {Bruel}, {Burnett}, {Busetto},
  {Caliandro}, {Cameron}, {Caraveo}, {Carius}, {Carlson}, {Casandjian},
  {Cavazzuti}, {Ceccanti}, {Cecchi}, {Charles}, {Chekhtman}, {Cheung},
  {Chiang}, {Chipaux}, {Cillis}, {Ciprini}, {Claus}, {Cohen-Tanugi},
  {Condamoor}, {Conrad}, {Corbet}, {Corucci}, {Costamante}, {Cutini}, {Davis},
  {Decotigny}, {DeKlotz}, {Dermer}, {de Angelis}, {Digel}, {do Couto e Silva},
  {Drell}, {Dubois}, {Dumora}, {Edmonds}, {Fabiani}, {Farnier}, {Favuzzi},
  {Flath}, {Fleury}, {Focke}, {Funk}, {Fusco}, {Gargano}, {Gasparrini},
  {Gehrels}, {Gentit}, {Germani}, {Giebels}, {Giglietto}, {Giommi}, {Giordano},
  {Glanzman}, {Godfrey}, {Grenier}, {Grondin}, {Grove}, {Guillemot}, {Guiriec},
  {Haller}, {Harding}, {Hart}, {Hays}, {Healey}, {Hirayama}, {Hjalmarsdotter},
  {Horn}, {Hughes}, {J{\'o}hannesson}, {Johansson}, {Johnson}, {Johnson},
  {Johnson}, {Johnson}, {Kamae}, {Katagiri}, {Kataoka}, {Kavelaars}, {Kawai},
  {Kelly}, {Kerr}, {Klamra}, {Kn{\"o}dlseder}, {Kocian}, {Komin}, {Kuehn},
  {Kuss}, {Landriu}, {Latronico}, {Lee}, {Lee}, {Lemoine-Goumard}, {Lionetto},
  {Longo}, {Loparco}, {Lott}, {Lovellette}, {Lubrano}, {Madejski}, {Makeev},
  {Marangelli}, {Massai}, {Mazziotta}, {McEnery}, {Menon}, {Meurer},
  {Michelson}, {Minuti}, {Mirizzi}, {Mitthumsiri}, {Mizuno}, {Moiseev},
  {Monte}, {Monzani}, {Moretti}, {Morselli}, {Moskalenko}, {Murgia},
  {Nakamori}, {Nishino}, {Nolan}, {Norris}, {Nuss}, {Ohno}, {Ohsugi}, {Omodei},
  {Orlando}, {Ormes}, {Paccagnella}, {Paneque}, {Panetta}, {Parent}, {Pearce},
  {Pepe}, {Perazzo}, {Pesce-Rollins}, {Picozza}, {Pieri}, {Pinchera}, {Piron},
  {Porter}, {Poupard}, {Rain{\`o}}, {Rando}, {Rapposelli}, {Razzano}, {Reimer},
  {Reimer}, {Reposeur}, {Reyes}, {Ritz}, {Rochester}, {Rodriguez}, {Romani},
  {Roth}, {Russell}, {Ryde}, {Sabatini}, {Sadrozinski}, {Sanchez}, {Sand er},
  {Sapozhnikov}, {Parkinson}, {Scargle}, {Schalk}, {Scolieri}, {Sgr{\`o}},
  {Share}, {Shaw}, {Shimokawabe}, {Shrader}, {Sierpowska-Bartosik}, {Siskind},
  {Smith}, {Smith}, {Spandre}, {Spinelli}, {Starck}, {Stephens}, {Strickman},
  {Strong}, {Suson}, {Tajima}, {Takahashi}, {Takahashi}, {Tanaka}, {Tenze},
  {Tether}, {Thayer}, {Thayer}, {Thompson}, {Tibaldo}, {Tibolla}, {Torres},
  {Tosti}, {Tramacere}, {Turri}, {Usher}, {Vilchez}, {Vitale}, {Wang},
  {Watters}, {Winer}, {Wood}, {Ylinen}, \& {Ziegler}}]{Atwood2009}
{Atwood}, W.~B., {Abdo}, A.~A., {Ackermann}, M., {et~al.} 2009, \apj, 697,
  1071, \dodoi{10.1088/0004-637X/697/2/1071}

\bibitem[{{Bai} \& {Spitkovsky}(2010)}]{Bai2010}
{Bai}, X.-N., \& {Spitkovsky}, A. 2010, \apj, 715, 1282,
  \dodoi{10.1088/0004-637X/715/2/1282}

\bibitem[{{Brambilla} {et~al.}(2015){Brambilla}, {Kalapotharakos}, {Harding},
  \& {Kazanas}}]{Brambilla2015}
{Brambilla}, G., {Kalapotharakos}, C., {Harding}, A.~K., \& {Kazanas}, D. 2015,
  \apj, 804, 84, \dodoi{10.1088/0004-637X/804/2/84}

\bibitem[{{Brambilla} {et~al.}(2018){Brambilla}, {Kalapotharakos}, {Timokhin},
  {Harding}, \& {Kazanas}}]{Brambilla2018}
{Brambilla}, G., {Kalapotharakos}, C., {Timokhin}, A.~N., {Harding}, A.~K., \&
  {Kazanas}, D. 2018, \apj, 858, 81, \dodoi{10.3847/1538-4357/aab3e1}

\bibitem[{{Bulik} {et~al.}(2000){Bulik}, {Rudak}, \& {Dyks}}]{Bulik2000}
{Bulik}, T., {Rudak}, B., \& {Dyks}, J. 2000, \mnras, 317, 97,
  \dodoi{10.1046/j.1365-8711.2000.03662.x}

\bibitem[{{Cerutti} {et~al.}(2016{\natexlab{a}}){Cerutti}, {Mortier}, \&
  {Philippov}}]{Cerutti2016current}
{Cerutti}, B., {Mortier}, J., \& {Philippov}, A.~A. 2016{\natexlab{a}}, \mnras,
  463, L89, \dodoi{10.1093/mnrasl/slw162}

\bibitem[{{Cerutti} {et~al.}(2020){Cerutti}, {Philippov}, \&
  {Dubus}}]{Cerutti2020}
{Cerutti}, B., {Philippov}, A., \& {Dubus}, G. 2020, arXiv e-prints,
  arXiv:2008.11462.
\newblock \doarXiv{2008.11462}

\bibitem[{{Cerutti} {et~al.}(2016{\natexlab{b}}){Cerutti}, {Philippov}, \&
  {Spitkovsky}}]{Cerutti2016}
{Cerutti}, B., {Philippov}, A.~A., \& {Spitkovsky}, A. 2016{\natexlab{b}},
  \mnras, 457, 2401, \dodoi{10.1093/mnras/stw124}

\bibitem[{{Cheng} {et~al.}(1986{\natexlab{a}}){Cheng}, {Ho}, \&
  {Ruderman}}]{Cheng1986I}
{Cheng}, K.~S., {Ho}, C., \& {Ruderman}, M. 1986{\natexlab{a}}, \apj, 300, 500,
  \dodoi{10.1086/163829}

\bibitem[{{Cheng} {et~al.}(1986{\natexlab{b}}){Cheng}, {Ho}, \&
  {Ruderman}}]{Cheng1986II}
---. 1986{\natexlab{b}}, \apj, 300, 522, \dodoi{10.1086/163830}

\bibitem[{{Cherenkov Telescope Array Consortium} {et~al.}(2019){Cherenkov
  Telescope Array Consortium}, {Acharya}, {Agudo}, {Al Samarai}, {Alfaro},
  {Alfaro}, {Alispach}, {Alves Batista}, {Amans}, {Amato}, {Ambrosi},
  {Antolini}, {Antonelli}, {Aramo}, {Araya}, {Armstrong}, {Arqueros},
  {Arrabito}, {Asano}, {Ashley}, {Backes}, {Balazs}, {Balbo}, {Ballester},
  {Ballet}, {Bamba}, {Barkov}, {Barres de Almeida}, {Barrio}, {Bastieri},
  {Becherini}, {Belfiore}, {Benbow}, {Berge}, {Bernardini}, {Bernardini},
  {Bernardos}, {Bernl{\"o}hr}, {Bertucci}, {Biasuzzi}, {Bigongiari}, {Biland},
  {Bissaldi}, {Biteau}, {Blanch}, {Blazek}, {Boisson}, {Bolmont}, {Bonanno},
  {Bonardi}, {Bonavolont{\`a}}, {Bonnoli}, {Bosnjak}, {B{\"o}ttcher},
  {Braiding}, {Bregeon}, {Brill}, {Brown}, {Brun}, {Brunetti}, {Buanes},
  {Buckley}, {Bugaev}, {B{\"u}hler}, {Bulgarelli}, {Bulik}, {Burton},
  {Burtovoi}, {Busetto}, {Canestrari}, {Capalbi}, {Capitanio}, {Caproni},
  {Caraveo}, {C{\'a}rdenas}, {Carlile}, {Carosi}, {Carqu{\'\i}n}, {Carr},
  {Casanova}, {Cascone}, {Catalani}, {Catalano}, {Cauz}, {Cerruti}, {Chadwick},
  {Chaty}, {Chaves}, {Chen}, {Chen}, {Chernyakova}, {Chikawa}, {Christov},
  {Chudoba}, {Cie{\'s}lar}, {Coco}, {Colafrancesco}, {Colin}, {Conforti},
  {Connaughton}, {Conrad}, {Contreras}, {Cortina}, {Costa}, {Costantini},
  {Cotter}, {Covino}, {Crocker}, {Cuadra}, {Cuevas}, {Cumani}, {D'A{\`\i}},
  {D'Ammando}, {D'Avanzo}, {D'Urso}, {Daniel}, {Davids}, {Dawson}, {Dazzi}, {De
  Angelis}, {de C{\'a}ssia dos Anjos}, {De Cesare}, {De Franco}, {de Gouveia
  Dal Pino}, {de la Calle}, {de los Reyes Lopez}, {De Lotto}, {De Luca}, {De
  Lucia}, {de Naurois}, {de O{\~n}a Wilhelmi}, {De Palma}, {De Persio}, {de
  Souza}, {Deil}, {Del Santo}, {Delgado}, {della Volpe}, {Di Girolamo}, {Di
  Pierro}, {Di Venere}, {D{\'\i}az}, {Dib}, {Diebold}, {Djannati-Ata{\"\i}},
  {Dom{\'\i}nguez}, {Dominis Prester}, {Dorner}, {Doro}, {Drass}, {Dravins},
  {Dubus}, {Dwarkadas}, {Ebr}, {Eckner}, {Egberts}, {Einecke}, {Ekoume},
  {Els{\"a}sser}, {Ernenwein}, {Espinoza}, {Evoli}, {Fairbairn},
  {Falceta-Goncalves}, {Falcone}, {Farnier}, {Fasola}, {Fedorova}, {Fegan},
  {Fernandez-Alonso}, {Fern{\'a}ndez-Barral}, {Ferrand}, {Fesquet},
  {Filipovic}, {Fioretti}, {Fontaine}, {Fornasa}, {Fortson}, {Freixas
  Coromina}, {Fruck}, {Fujita}, {Fukazawa}, {Funk}, {F{\"u}{\ss}ling},
  {Gabici}, {Gadola}, {Gallant}, {Garcia}, {Garcia L{\'o}pez}, {Garczarczyk},
  {Gaskins}, {Gasparetto}, {Gaug}, {Gerard}, {Giavitto}, {Giglietto}, {Giommi},
  {Giordano}, {Giro}, {Giroletti}, {Giuliani}, {Glicenstein}, {Gnatyk},
  {Godinovic}, {Goldoni}, {G{\'o}mez-Vargas}, {Gonz{\'a}lez}, {Gonz{\'a}lez},
  {G{\"o}tz}, {Graham}, {Grandi}, {Granot}, {Green}, {Greenshaw}, {Griffiths},
  {Gunji}, {Hadasch}, {Hara}, {Hardcastle}, {Hassan}, {Hayashi}, {Hayashida},
  {Heller}, {Helo}, {Hermann}, {Hinton}, {Hnatyk}, {Hofmann}, {Holder},
  {Horan}, {H{\"o}randel}, {Horns}, {Horvath}, {Hovatta}, {Hrabovsky},
  {Hrupec}, {Humensky}, {H{\"u}tten}, {Iarlori}, {Inada}, {Inome}, {Inoue},
  {Inoue}, {Inoue}, {Iocco}, {Ioka}, {Iori}, {Ishio}, {Iwamura}, {Jamrozy},
  {Janecek}, {Jankowsky}, {Jean}, {Jung-Richardt}, {Jurysek}, {Kaaret},
  {Karkar}, {Katagiri}, {Katz}, {Kawanaka}, {Kazanas}, {Kh{\'e}lifi}, {Kieda},
  {Kimeswenger}, {Kimura}, {Kisaka}, {Knapp}, {Kn{\"o}dlseder}, {Koch},
  {Kohri}, {Komin}, {Kosack}, {Kraus}, {Krause}, {Krau{\ss}}, {Kubo}, {Kukec
  Mezek}, {Kuroda}, {Kushida}, {La Palombara}, {Lamanna}, {Lang}, {Lapington},
  {Le Blanc}, {Leach}, {Lees}, {Lefaucheur}, {Leigui de Oliveira}, {Lenain},
  {Lico}, {Limon}, {Lindfors}, {Lohse}, {Lombardi}, {Longo}, {L{\'o}pez},
  {L{\'o}pez-Coto}, {Lu}, {Lucarelli}, {Luque-Escamilla}, {Lyard}, {Maccarone},
  {Maier}, {Majumdar}, {Malaguti}, {Mandat}, {Maneva}, {Manganaro}, {Mangano},
  {Marcowith}, {Mar{\'\i}n}, {Markoff}, {Mart{\'\i}}, {Martin},
  {Mart{\'\i}nez}, {Mart{\'\i}nez}, {Masetti}, {Masuda}, {Maurin}, {Maxted},
  {Mazin}, {Medina}, {Melandri}, {Mereghetti}, {Meyer}, {Minaya}, {Mirabal},
  {Mirzoyan}, {Mitchell}, {Mizuno}, {Moderski}, {Mohammed}, {Mohrmann},
  {Montaruli}, {Moralejo}, {Morcuende-Parrilla}, {Mori}, {Morlino}, {Morris},
  {Morselli}, {Moulin}, {Mukherjee}, {Mundell}, {Murach}, {Muraishi}, {Murase},
  {Nagai}, {Nagataki}, {Nagayoshi}, {Naito}, {Nakamori}, {Nakamura}, {Niemiec},
  {Nieto}, {Niko{\l}ajuk}, {Nishijima}, {Noda}, {Nosek}, {Novosyadlyj},
  {Nozaki}, {O'Brien}, {Oakes}, {Ohira}, {Ohishi}, {Ohm}, {Okazaki}, {Okumura},
  {Ong}, {Orienti}, {Orito}, {Osborne}, {Ostrowski}, {Otte}, {Oya}, {Padovani},
  {Paizis}, {Palatiello}, {Palatka}, {Paoletti}, {Paredes}, {Pareschi},
  {Parsons}, {Pe'er}, {Pech}, {Pedaletti}, {Perri}, {Persic}, {Petrashyk},
  {Petrucci}, {Petruk}, {Peyaud}, {Pfeifer}, {Piano}, {Pisarski}, {Pita},
  {Pohl}, {Polo}, {Pozo}, {Prandini}, {Prast}, {Principe}, {Prokhorov},
  {Prokoph}, {Prouza}, {P{\"u}hlhofer}, {Punch}, {P{\"u}rckhauer}, {Queiroz},
  {Quirrenbach}, {Rain{\`o}}, {Razzaque}, {Reimer}, {Reimer}, {Reisenegger},
  {Renaud}, {Rezaeian}, {Rhode}, {Ribeiro}, {Rib{\'o}}, {Richtler}, {Rico},
  {Rieger}, {Riquelme}, {Rivoire}, {Rizi}, {Rodriguez}, {Rodriguez Fernandez},
  {Rodr{\'\i}guez V{\'a}zquez}, {Rojas}, {Romano}, {Romeo}, {Rosado}, {Rovero},
  {Rowell}, {Rudak}, {Rugliancich}, {Rulten}, {Sadeh}, {Safi-Harb}, {Saito},
  {Sakaki}, {Sakurai}, {Salina}, {S{\'a}nchez-Conde}, {Sandaker}, {Sandoval},
  {Sangiorgi}, {Sanguillon}, {Sano}, {Santander}, {Sarkar}, {Satalecka},
  {Saturni}, {Schioppa}, {Schlenstedt}, {Schneider}, {Schoorlemmer},
  {Schovanek}, {Schulz}, {Schussler}, {Schwanke}, {Sciacca}, {Scuderi},
  {Seitenzahl}, {Semikoz}, {Sergijenko}, {Servillat}, {Shalchi}, {Shellard},
  {Sidoli}, {Siejkowski}, {Sillanp{\"a}{\"a}}, {Sironi}, {Sitarek}, {Sliusar},
  {Slowikowska}, {Sol}, {Stamerra}, {Stani{\v{c}}}, {Starling}, {Stawarz},
  {Stefanik}, {Stephan}, {Stolarczyk}, {Stratta}, {Straumann}, {Suomijarvi},
  {Supanitsky}, {Tagliaferri}, {Tajima}, {Tavani}, {Tavecchio}, {Tavernet},
  {Tayabaly}, {Tejedor}, {Temnikov}, {Terada}, {Terrier}, {Terzic}, {Teshima},
  {Testa}, {Thoudam}, {Tian}, {Tibaldo}, {Tluczykont}, {Todero Peixoto},
  {Tokanai}, {Tomastik}, {Tonev}, {Tornikoski}, {Torres}, {Torresi}, {Tosti},
  {Tothill}, {Tovmassian}, {Travnicek}, {Trichard}, {Trifoglio}, {Troyano
  Pujadas}, {Tsujimoto}, {Umana}, {Vagelli}, {Vagnetti}, {Valentino},
  {Vallania}, {Valore}, {van Eldik}, {Vandenbroucke}, {Varner}, {Vasileiadis},
  {Vassiliev}, {V{\'a}zquez Acosta}, {Vecchi}, {Vega}, {Vercellone}, {Veres},
  {Vergani}, {Verzi}, {Vettolani}, {Viana}, {Vigorito}, {Villanueva}, {Voelk},
  {Vollhardt}, {Vorobiov}, {Vrastil}, {Vuillaume}, {Wagner}, {Wagner},
  {Walter}, {Ward}, {Warren}, {Watson}, {Werner}, {White}, {White},
  {Wierzcholska}, {Wilcox}, {Will}, {Williams}, {Wischnewski}, {Wood},
  {Yamamoto}, {Yamazaki}, {Yanagita}, {Yang}, {Yoshida}, {Yoshiike},
  {Yoshikoshi}, {Zacharias}, {Zaharijas}, {Zampieri}, {Zandanel}, {Zanin},
  {Zavrtanik}, {Zavrtanik}, {Zdziarski}, {Zech}, {Zechlin}, {Zhdanov},
  {Ziegler}, \& {Zorn}}]{Acharya2019}
{Cherenkov Telescope Array Consortium}, {Acharya}, B.~S., {Agudo}, I., {et~al.}
  2019, {Science with the Cherenkov Telescope Array}, \dodoi{10.1142/10986}

\bibitem[{{Contopoulos} \& {Kalapotharakos}(2010)}]{Contopoulos2010}
{Contopoulos}, I., \& {Kalapotharakos}, C. 2010, 404, 767,
  \dodoi{10.1111/j.1365-2966.2010.16338.x}

\bibitem[{{Daugherty} \& {Harding}(1982)}]{Daugherty1982}
{Daugherty}, J.~K., \& {Harding}, A.~K. 1982, \apj, 252, 337,
  \dodoi{10.1086/159561}

\bibitem[{{Deutsch}(1955)}]{Deutsch1955}
{Deutsch}, A.~J. 1955, Annales d'Astrophysique, 18, 1

\bibitem[{{Dyks} {et~al.}(2004){Dyks}, {Harding}, \& {Rudak}}]{Dyks2004}
{Dyks}, J., {Harding}, A.~K., \& {Rudak}, B. 2004, \apj, 606, 1125,
  \dodoi{10.1086/383121}

\bibitem[{{Fr{a}ckowiak} \& {Rudak}(2005)}]{Frackowiak2005}
{Fr{a}ckowiak}, M., \& {Rudak}, B. 2005, Advances in Space Research, 35, 1152,
  \dodoi{10.1016/j.asr.2005.02.031}

\bibitem[{{Gotthelf} \& {Bogdanov}(2017)}]{Gotthelf2017}
{Gotthelf}, E.~V., \& {Bogdanov}, S. 2017, \apj, 845, 159,
  \dodoi{10.3847/1538-4357/aa813c}

\bibitem[{{Gotthelf} {et~al.}(2002){Gotthelf}, {Halpern}, \&
  {Dodson}}]{Gotthelf2002}
{Gotthelf}, E.~V., {Halpern}, J.~P., \& {Dodson}, R. 2002, \apjl, 567, L125,
  \dodoi{10.1086/340109}

\bibitem[{{Harding} \& {Kalapotharakos}(2015)}]{Harding2015}
{Harding}, A.~K., \& {Kalapotharakos}, C. 2015, \apj, 811, 63,
  \dodoi{10.1088/0004-637X/811/1/63}

\bibitem[{{Harding} {et~al.}(2018){Harding}, {Kalapotharakos}, {Barnard}, \&
  {Venter}}]{Harding2018}
{Harding}, A.~K., {Kalapotharakos}, C., {Barnard}, M., \& {Venter}, C. 2018,
  \apjl, 869, L18, \dodoi{10.3847/2041-8213/aaf3b2}

\bibitem[{{Harding} \& {Muslimov}(1998)}]{Harding1998}
{Harding}, A.~K., \& {Muslimov}, A.~G. 1998, \apj, 508, 328,
  \dodoi{10.1086/306394}

\bibitem[{{Harding} \& {Muslimov}(2011)}]{Harding2011}
---. 2011, \apj, 743, 181, \dodoi{10.1088/0004-637X/743/2/181}

\bibitem[{{Harding} {et~al.}(2002{\natexlab{a}}){Harding}, {Muslimov}, \&
  {Zhang}}]{Harding2002a}
{Harding}, A.~K., {Muslimov}, A.~G., \& {Zhang}, B. 2002{\natexlab{a}}, \apj,
  576, 366, \dodoi{10.1086/341633}

\bibitem[{{Harding} {et~al.}(2008){Harding}, {Stern}, {Dyks}, \&
  {Frackowiak}}]{Harding2008}
{Harding}, A.~K., {Stern}, J.~V., {Dyks}, J., \& {Frackowiak}, M. 2008, \apj,
  680, 1378, \dodoi{10.1086/588037}

\bibitem[{{Harding} {et~al.}(2002{\natexlab{b}}){Harding}, {Strickman},
  {Gwinn}, {Dodson}, {Moffet}, \& {McCulloch}}]{Harding2002}
{Harding}, A.~K., {Strickman}, M.~S., {Gwinn}, C., {et~al.} 2002{\natexlab{b}},
  \apj, 576, 376, \dodoi{10.1086/341732}

\bibitem[{{Harding} {et~al.}(2005){Harding}, {Usov}, \&
  {Muslimov}}]{Harding2005}
{Harding}, A.~K., {Usov}, V.~V., \& {Muslimov}, A.~G. 2005, \apj, 622, 531,
  \dodoi{10.1086/427840}

\bibitem[{{Hirotani}(2001)}]{Hirotani2001}
{Hirotani}, K. 2001, \apj, 549, 495, \dodoi{10.1086/319038}

\bibitem[{{Holler} {et~al.}(2015){Holler}, {Berge}, {van Eldik}, {Lenain},
  {Marandon}, {Murach}, {de Naurois}, {Parsons}, {Prokoph}, \&
  {Zaborov}}]{Holler2015}
{Holler}, M., {Berge}, D., {van Eldik}, C., {et~al.} 2015, arXiv e-prints,
  arXiv:1509.02902.
\newblock \doarXiv{1509.02902}

\bibitem[{{Kalapotharakos} {et~al.}(2018){Kalapotharakos}, {Brambilla},
  {Timokhin}, {Harding}, \& {Kazanas}}]{Kalapotharakos2018}
{Kalapotharakos}, C., {Brambilla}, G., {Timokhin}, A., {Harding}, A.~K., \&
  {Kazanas}, D. 2018, \apj, 857, 44, \dodoi{10.3847/1538-4357/aab550}

\bibitem[{{Kalapotharakos} \& {Contopoulos}(2009)}]{Kalapotharakos2009}
{Kalapotharakos}, C., \& {Contopoulos}, I. 2009, \aap, 496, 495,
  \dodoi{10.1051/0004-6361:200810281}

\bibitem[{{Kalapotharakos} {et~al.}(2014){Kalapotharakos}, {Harding}, \&
  {Kazanas}}]{Kalapotharakos2014}
{Kalapotharakos}, C., {Harding}, A.~K., \& {Kazanas}, D. 2014, \apj, 793, 97,
  \dodoi{10.1088/0004-637X/793/2/97}

\bibitem[{{Kalapotharakos} {et~al.}(2017){Kalapotharakos}, {Harding},
  {Kazanas}, \& {Brambilla}}]{Kalapotharakos2017}
{Kalapotharakos}, C., {Harding}, A.~K., {Kazanas}, D., \& {Brambilla}, G. 2017,
  \apj, 842, 80, \dodoi{10.3847/1538-4357/aa713a}

\bibitem[{{Kalapotharakos} {et~al.}(2012){Kalapotharakos}, {Kazanas},
  {Harding}, \& {Contopoulos}}]{Kalapotharakos2012}
{Kalapotharakos}, C., {Kazanas}, D., {Harding}, A., \& {Contopoulos}, I. 2012,
  \apj, 749, 2, \dodoi{10.1088/0004-637X/749/1/2}

\bibitem[{{Kargaltsev} {et~al.}(2005){Kargaltsev}, {Pavlov}, {Zavlin}, \&
  {Romani}}]{Kargaltsev2005}
{Kargaltsev}, O.~Y., {Pavlov}, G.~G., {Zavlin}, V.~E., \& {Romani}, R.~W. 2005,
  \apj, 625, 307, \dodoi{10.1086/429368}

\bibitem[{{Kuiper} {et~al.}(1996){Kuiper}, {Hermsen}, {Bennett}, {Connors},
  {Much}, {Ryan}, {Schoenfelder}, \& {Strong}}]{Kuiper1996}
{Kuiper}, L., {Hermsen}, W., {Bennett}, K., {et~al.} 1996, \aaps, 120, 73

\bibitem[{{Kuiper} {et~al.}(2001){Kuiper}, {Hermsen}, {Cusumano}, {Diehl},
  {Sch{\"o}nfelder}, {Strong}, {Bennett}, \& {McConnell}}]{Kuiper2001}
{Kuiper}, L., {Hermsen}, W., {Cusumano}, G., {et~al.} 2001, \aap, 378, 918,
  \dodoi{10.1051/0004-6361:20011256}

\bibitem[{{Li} {et~al.}(2012){Li}, {Spitkovsky}, \& {Tchekhovskoy}}]{Li2012}
{Li}, J., {Spitkovsky}, A., \& {Tchekhovskoy}, A. 2012, \apj, 746, 60,
  \dodoi{10.1088/0004-637X/746/1/60}

\bibitem[{{McCann}(2015)}]{McCann15}
{McCann}, A. 2015, arXiv e-prints, arXiv:1503.01670.
\newblock \doarXiv{1503.01670}

\bibitem[{{Ng} \& {Romani}(2008)}]{Ng2008}
{Ng}, C.~Y., \& {Romani}, R.~W. 2008, \apj, 673, 411, \dodoi{10.1086/523935}

\bibitem[{{Philippov} {et~al.}(2019){Philippov}, {Uzdensky}, {Spitkovsky}, \&
  {Cerutti}}]{Philippov2019}
{Philippov}, A., {Uzdensky}, D.~A., {Spitkovsky}, A., \& {Cerutti}, B. 2019,
  \apjl, 876, L6, \dodoi{10.3847/2041-8213/ab1590}

\bibitem[{{Philippov} \& {Spitkovsky}(2018)}]{Philippov2018}
{Philippov}, A.~A., \& {Spitkovsky}, A. 2018, \apj, 855, 94,
  \dodoi{10.3847/1538-4357/aaabbc}

\bibitem[{{Romani}(1996)}]{Romani1996}
{Romani}, R.~W. 1996, \apj, 470, 469, \dodoi{10.1086/177878}

\bibitem[{{Romani} \& {Yadigaroglu}(1995)}]{Romani1995}
{Romani}, R.~W., \& {Yadigaroglu}, I.~A. 1995, \apj, 438, 314,
  \dodoi{10.1086/175076}

\bibitem[{{Rudak} \& {Dyks}(2017)}]{Rudak2017}
{Rudak}, B., \& {Dyks}, J. 2017, in Proceedings of the 7th International Fermi
  Symposium, 15

\bibitem[{{Shibanov} {et~al.}(2003){Shibanov}, {Koptsevich}, {Sollerman}, \&
  {Lundqvist}}]{Shibanov2003}
{Shibanov}, Y.~A., {Koptsevich}, A.~B., {Sollerman}, J., \& {Lundqvist}, P.
  2003, \aap, 406, 645, \dodoi{10.1051/0004-6361:20030652}

\bibitem[{{Shibanov} {et~al.}(2006){Shibanov}, {Zharikov}, {Komarova}, {Kawai},
  {Urata}, {Koptsevich}, {Sokolov}, {Shibata}, \& {Shibazaki}}]{Shibanov2006}
{Shibanov}, Y.~A., {Zharikov}, S.~V., {Komarova}, V.~N., {et~al.} 2006, \aap,
  448, 313, \dodoi{10.1051/0004-6361:20054178}

\bibitem[{{Sollerman} {et~al.}(2019){Sollerman}, {Selsing}, {Vreeswijk},
  {Lundqvist}, \& {Nyholm}}]{Sollerman2019}
{Sollerman}, J., {Selsing}, J., {Vreeswijk}, P.~M., {Lundqvist}, P., \&
  {Nyholm}, A. 2019, \aap, 629, A140, \dodoi{10.1051/0004-6361/201935086}

\bibitem[{{Spir-Jacob} {et~al.}(2019){Spir-Jacob}, {Djannati-Ata{\"\i}},
  {Mohrmann}, {Giavitto}, {Kh{\'e}lifi}, {Rudak}, {Venter}, \&
  {Zanin}}]{SpirJacob2019}
{Spir-Jacob}, M., {Djannati-Ata{\"\i}}, A., {Mohrmann}, L., {et~al.} 2019,
  arXiv e-prints, arXiv:1908.06464.
\newblock \doarXiv{1908.06464}

\bibitem[{{Takata} {et~al.}(2006){Takata}, {Shibata}, {Hirotani}, \&
  {Chang}}]{Takata2006}
{Takata}, J., {Shibata}, S., {Hirotani}, K., \& {Chang}, H.~K. 2006, \mnras,
  366, 1310, \dodoi{10.1111/j.1365-2966.2006.09904.x}

\bibitem[{{Tang} {et~al.}(2008){Tang}, {Takata}, {Jia}, \& {Cheng}}]{Tang2008}
{Tang}, A. P.~S., {Takata}, J., {Jia}, J.~J., \& {Cheng}, K.~S. 2008, \apj,
  676, 562, \dodoi{10.1086/527029}

\bibitem[{{Timokhin}(2010)}]{Timokhin2010}
{Timokhin}, A.~N. 2010, \mnras, 408, 2092,
  \dodoi{10.1111/j.1365-2966.2010.17286.x}

\bibitem[{{Timokhin} \& {Arons}(2013)}]{Timokhin2013}
{Timokhin}, A.~N., \& {Arons}, J. 2013, \mnras, 429, 20,
  \dodoi{10.1093/mnras/sts298}

\bibitem[{{Timokhin} \& {Harding}(2015)}]{Timokhin2015}
{Timokhin}, A.~N., \& {Harding}, A.~K. 2015, \apj, 810, 144,
  \dodoi{10.1088/0004-637X/810/2/144}

\bibitem[{{Timokhin} \& {Harding}(2019)}]{Timokhin2019}
---. 2019, \apj, 871, 12, \dodoi{10.3847/1538-4357/aaf050}

\bibitem[{{Torres}(2018)}]{Torres2018}
{Torres}, D.~F. 2018, Nature Astronomy, 2, 247,
  \dodoi{10.1038/s41550-018-0384-5}

\bibitem[{{Venter} \& {De Jager}(2005)}]{Venter2005}
{Venter}, C., \& {De Jager}, O.~C. 2005, \apjl, 619, L167,
  \dodoi{10.1086/427991}

\end{thebibliography}
\bibliographystyle{aasjournal}

\newpage

\newpage

\newpage

\newpage

\newpage

\newpage

\newpage

\newpage

\newpage

\newpage

\newpage

\newpage

\newpage

\newpage

\end{document}